\documentclass{llncs}
\title{Almost 2-SAT is Fixed-Parameter Tractable}
\author{Igor Razgon and Barry O'Sullivan\\
\{i.razgon,b.osullivan\}@cs.ucc.ie }
\institute{Computer Science
Department, University College Cork, Ireland}

\newtheorem{cclaim}{Claim}
\usepackage{amssymb}
\usepackage{url}

\begin{document}
\maketitle
\begin{abstract}
We consider the following problem. Given a 2-CNF formula, is it
possible to remove at most $k$ clauses so that the resulting 2-CNF
formula is satisfiable? This problem is known to different research
communities in Theoretical Computer Science under the names 'Almost
2-SAT', 'All-but-$k$ 2-SAT', '2-CNF deletion', '2-SAT deletion'. The
status of fixed-parameter tractability of this problem is a
long-standing open question in the area of Parameterized Complexity.
We resolve this open question by proposing an algorithm which solves
this problem in $O(15^k*k*m^3)$ and thus we show that this problem
is fixed-parameter tractable.
\end{abstract}

\section{Introduction}
We consider the following problem. Given a \textsc{2-cnf} formula,
is it possible to remove at most $k$ clauses so that the resulting
\textsc{2-cnf} formula is satisfiable? This problem is known to
different research communities in Theoretical Computer Science under
the names 'Almost 2-SAT', 'All-but-$k$ 2-SAT', '\textsc{2-cnf}
deletion', '2-SAT deletion'. The status of fixed-parameter
tractability of this problem is a long-standing open question in the
area of Parameterized Complexity. The question regarding the
fixed-parameter tractability of this problem was first raised in
1997 by Mahajan and Raman \cite{RamanECCC} (see \cite{RamanJALG} for
the journal version). This question has been posed in the book of
Niedermeier \cite{Niederbook} being referred as one of central
challenges for parameterized algorithms design. Finally, in July
2007, this question was included by Fellows in the list of open
problems of the Dagstuhl seminar on Parameterized Complexity
\cite{dagopen}. In this paper we resolve this open question by
proposing an algorithm that solves this problem in $O(15^k*k*m^3)$
time. Thus we show that this problem is fixed-parameter tractable
(\textsc{fpt}).

\subsection{Overview of the algorithm}
We start from the terminology we adopt regarding the names of the
considered problems. We call \emph{Almost 2-SAT} (abbreviated as
\emph{2-ASAT}) the optimization problem whose output is the smallest
subset of clauses that have to be removed from the given 2-CNF
formula so that the resulting 2-CNF formula is satisfiable. The
\emph{parameterized 2-ASAT} problem gets as additional input a
parameter $k$ and the output of this problem is a set of at most $k$
clauses whose removal makes the given 2-CNF formula satisfiable, in
case such a set exists. If there is no such a set, the output is
'NO'. So, the algorithm proposed in this paper solves the
parameterized 2-ASAT problem.

We introduce a variation of the 2-ASAT problem called the
\emph{annotated 2-ASAT problem with a single literal} abbreviated as
\emph{2-ASLASAT}. The input of this problem is $(F,L,l)$, where $F$
is a 2-CNF formula, $L$ is a set of literals such that $F$ is
\emph{satisfiable w.r.t. $L$} (i.e. has a satisfying assignment
which does not include negations of literals of $L$), $l$ is a
single literal. The task is to find a smallest subset of clauses of
$F$ such that after their removal the resulting formula is
satisfiable w.r.t. $(L \cup \{l\})$. The parameterized versions of
the 2-ASLASAT problem is defined analogously to the parameterized
2-ASAT problem.

The description of the algorithm for the parameterized 2-ASAT
problem is divided into two parts. In the first part (which is the
most important one) we provide an algorithm which solves the
parameterized 2-ASLASAT problem in $O^*(5^k)$ time. In the second
part we show that the parameterized 2-ASAT problem can be solved by
$O^*(3^k)$ applications of the algorithm solving the parameterized
2-ASLASAT problem. The resulting runtime follows from the product of
the last two complexity expressions. The transformation of the
2-ASAT problem into the 2-ASLASAT problem is based on the
\emph{iterative compression} and can be seen as an adaptation of the
method employed in \cite{HuffnerWEA} in order to solve the graph
bipartization problem. In the rest of the subsection we overview the
first part.

In order to show that the 2-ASLASAT problem is FPT, we represent the
2-ASLASAT problem as a \emph{separation} problem and prove a number
of theorems based on this view. In particular, we introduce a notion
of a \emph{walk} from a literal $l'$ to a literal $l''$ in a 2-CNF
formula $F$. We define the walk as a sequence $(l' \vee l_1),(\neg
l_1 \vee l_2), \dots, (\neg l_{k-1} \vee l_k),(\neg l_k \vee l'')$
of clauses of $F$ such that literals are ordered within each clause
so that the \emph{second} literal of each clause except the last one
is the negation of the \emph{first} literal of the next clause. Then
we prove that, given an instance $(F,L,l)$ of the 2-ASLASAT problem,
$F$ is insatisfiable w.r.t. $L \cup \{l\}$ if and only if there is a
walk from $\neg L$ (i.e. from the set of negations of the literals
of $L$) to $\neg l$ or a walk from $\neg l$ to $\neg l$. Thus the
2-ASLASAT problem can be viewed as a problem of finding the smallest
set of clauses whose removal breaks all these walks.

Next we define the notion of a \emph{path} of $F$ as a walk of $F$
with no repeated clauses. Based on this notion we prove a Menger's
like theorem. In particular, given an instance $(F,L,l)$ of the
2-ASLASAT problem, we show that the smallest number of clauses whose
removal breaks all the paths from $\neg L$ to $\neg l$ equals the
largest number of clause-disjoint paths from $\neg L$ to $\neg l$
(for this result it is essential that $F$ is satisfiable w.r.t.
$L$). Based on this result, we show that the size of the above
\emph{smallest separator} of $\neg L$ from from $\neg l$ can be
computed in a polynomial time by a Ford-Fulkerson-like procedure.
Thus this size is a polynomially computable \emph{lower bound} on
the size of the solution of $(F,L,l)$.

Next we introduce the notion of a \emph{neutral literal} $l^*$ of
$(F,L,l)$ whose main property is that the number of clauses which
separate $\neg (L \cup \{l^*\})$ from $\neg l$ equals the number of
clauses separating $\neg L$ from $\neg l$. Then we prove a theorem
stating that in this case the size of a solution of $(F,L \cup
\{l^*\},l)$ does not exceed the size of a solution of $(F,L,l)$. The
strategy of the proof is similar to the strategy of the proof of the
main theorem of \cite{ChenWADS2007}.

Having proved all the above theorems, we present the algorithm
solving the parameterized 2-ASLASAT problem on input $(F,L,l,k)$.
The algorithm selects a clause $C$. If $C$ includes a neutral
literal $l^*$ then the algorithm applies itself recursively to $(F,L
\cup \{l^*\},l,k)$ (this operation is justified by the theorem in
the previous paragraph). If not, the algorithm produces at most
three branches on one of them it removes $C$ from $F$ and decreases
the parameter. On each of the other branches the algorithm adds one
of literals of $C$ to $L$ and applies itself recursively without
changing the size of the parameter. The search tree produced by the
algorithm is bounded because on each branch either the parameter is
decreased or the lower bound on the solution size is increased
(because the literals of the selected clause are \emph{not
neutral}). Thus on each branch \emph{the gap between the parameter
and the lower bound of the solution size is decreased} which ensures
that the size of the search tree exponentially depends only on $k$
and not on the size of $F$.

\subsection{Related Work}
As said above, the parameterized \textsc{2-asat} problem has been
introduced in \cite{RamanECCC}. In \cite{KhotRaman00}, this problem
was shown to be a generalization of the parameterized graph
bipartization problem, which was also an open problem at that time.
The latter problem has been resolved in \cite{Reed1}. The additional
contribution of \cite{Reed1} was introducing a method of iterative
compression which has had a considerable impact on the design of
parameterized algorithms. The most recent algorithms based on this
method are currently the best algorithm for the undirected Feedback
Vertex Set \cite{ChenWADSFVS} and the first parameterized algorithm
for the famous Direct FVS problem \cite{ChenSTOC}. For earlier
results based on the iterative compression, we refer the reader to a
survey article \cite{HuffnerSurvey}.

The study of parameterized graph separation problems has been
initiated in \cite{MarxTCS}. The technique introduced by the author
allowed him to design fixed-parameter algorithms for the
multiterminal cut problem and for a more general multicut problem,
the latter assumed that the number of pairs of terminals to be
separated was also a parameter. The latter result has been extended
in \cite{GuoSOFFSEM} where fixed-parameter algorithms for multicut
problems on several classes of graphs have been proposed. The first
$O(c^k*poly(n))$ algorithm for the multiterminal cut problem has
been proposed in \cite{ChenWADS2007}. A reformulation of the main
theorem of \cite{ChenWADS2007} is an essential part of the
parameterized algorithm for the Directed FVS problem \cite{ChenSTOC}
mentioned in the previous paragraph. In the present paper, we
applied the strategy of proof of this theorem in order to show that
adding a \emph{neutral} literal to the set of literals of the input
does not increase the solution size. Along with computing the
separators, the methods of computing disjoint paths have been
investigated. The research led to intractability results
\cite{Slivkins1} and parameterized approximability results
\cite{GroheICALP}.

The parameterized MAX-SAT problem (a complementary problem to the
one considered in the present paper) where the goal is to satisfy at
least $k$ clauses of arbitrary sizes received a considerable
attention from the researchers resulted in a series of improvements
of the worst-case upper bound on the runtime of this problem.
Currently the best algorithm is given in \cite{ChenKanj2004} and
solves this problem in $O(1.37^k+|F|)$, where $|F|$ is the size of
the given formula.

\subsection{Structure of the Paper}
In Section 2 we introduce the terminology which we use in the rest
of the paper. In Section 3 we prove the theorems mentioned in the
above overview subsection. In Section 4 we present an algorithm for
the parameterized 2-ASLASAT problem, prove its correctness and
evaluate the runtime. In Section 5 we present the iterative
compression based transformation from parameterized 2-ASAT problem
to the parameterized 2-SLASAT problem.
\section{Terminology}

\subsection{2-CNF Formulas}
A CNF formula $F$ is called a \emph{2-CNF formula} if each clause of
$F$ is of size at most 2. Throughout the paper we make two
assumptions regarding the considered 2-CNF formulas. First, we
assume that all the clauses of the considered formulas are of size
2. If a formula has a clause $(l)$ of size 1 then this clause is
represented as $(l \vee l)$. Second, everywhere except the very last
theorem, we assume that all the clauses of any considered formula
are pairwise distinct. \footnote{Note that the clause $(l_1 \vee
l_2)$ is \emph{the same} as the clause $(l_2 \vee l_1)$.} This
assumption allows us to represent the operation of removal clauses
from a formula in a set-theoretical manner. In particular, let $S$
be a set of clauses \footnote{We implicitly assume that all the
clauses considered in this paper have size 2}. Then $F \setminus S$
is a 2-CNF formula which is the $AND$ of clauses of $F$ that are not
contained in $S$. The result of removal a single clause $C$ is
denoted by $F \setminus C$ rather than $F \setminus \{C\}$.

Let $F$, $S$, $C$, $L$ be a 2-CNF formula, a set of clauses, a
single clause, and a set of literals. Then $Var(F)$, $Var(S)$,
$Var(C)$, $Var(L)$ denote the set of variables whose literals appear
in $F$, $S$, $C$, and $L$, respectively. For a single literal $l$,
we denote by $Var(l)$ the variable of $l$. Also we denote by
$Clauses(F)$ the set of clauses of $F$. %Introduce $Literals(F)$ where it appears

A set of literals $L$ is called \emph{non-contradictory} if it does
not contain a literal and its negation. A literal $l$
\emph{satisfies} a clause $(l_1 \vee l_2)$ if $l=l_1$ or $l=l_2$.
Given a 2-CNF formula $F$, a non-contradictory set of literals $L$
such that $Var(F)=Var(L)$ and each clause of $F$ is satisfied by at
least one literal of $L$, we call $L$ a \emph{satisfying assignment}
of $F$. $F$ is \emph{satisfiable} if it has at least one satisfying
assignment. Given a set of literals $L$, we denote by $\neg L$ the
set consisting of negations of all the literals of $L$. For example,
if $L=\{l_1,l_2,\neg l_3\}$ then $\neg L=\{\neg l_1,\neg l_2,l_3\}$.

Let $F$ be a 2-CNF formula and $L$ be a set of literals. $F$ is
\emph{satisfiable with respect to} $L$ if $F$ has a satisfying
assignment $P$ which does not intersect with $\neg L$ \footnote{We
do not say '$P$ contains $L$' because generally $Var(L)$ may be not
a subset of $Var(F)$}. The notion of satisfiability of a 2-CNF
formula with respect to the given set of literals will be very
frequently used in the paper, hence, in order to save the space, we
introduce a special notation for this notion. In particular, we say
that $SWRT(F,L)$ is true (false) if $F$ is, respectively,
satisfiable (not satisfiable) with respect to $L$. If $L$ consists
of a single literal $l$ then we write $SWRT(F,l)$ rather than
$SWRT(F,\{l\})$.

\subsection{Walks and paths}
\begin{definition}
A walk of the given 2-CNF formula $F$ is a non-empty sequence
$w=(C_1, \dots, C_q)$ of (not necessarily distinct) clauses of $F$
having the following property. For each $C_i$ one of its literals is
specified as the \emph{first literal} of $C_i$, the other literal is
the \emph{second literal}, and for any two consecutive clauses $C_i$
and $C_{i+1}$ the second literal of $C_i$ is the negation of the
first literal of $C_{i+1}$.
\end{definition}

Let $w=(C_1, \dots, C_q)$ be a walk and let $l'$ and $l''$ be the
first literal of $C_1$ and the second literal of $C_q$,
respectively. Then we say that $l'$ is \emph{the first literal of}
$w$, that $l''$ is \emph{the last literal of} $w$, and that $w$ is
\emph{a walk from} $l'$ \emph{to} $l''$. Let $L$ be a set of
literals such that $l' \in L$. Then we say that $w$ is a walk
\emph{from} $L$. Let $C=(l_1 \vee l_2)$ be a clause of $w$. Then
$l_1$ is a first literal of $C$ with respect to (w.r.t.) $w$ if
$l_1$ is the first literal of some $C_i$ such that $C=C_i$. A second
literal of a clause with respect to a walk is defined accordingly.
(Generally a literal of a clause may be both a first and a second
with respect to the given walk, which is shown in the example
below). We denote by $reverse(w)$ a walk $(C_q, \dots, C_1)$ in
which the first and the second literals of each entry are exchanged
w.r.t. $w$. Given a clause $C''=(\neg l'' \vee l^*)$, we denote by
$w+(\neg l'' \vee l^*)$ the walk obtained by appending $C''$ to the
end of $w$ and setting $\neg l''$ to be the first literal of the
last entry of $w+(\neg l'' \vee l^*)$ and $l^*$ to be the second
one. More generally, let $w'$ be a walk whose first literal is $\neg
l''$. Then $w+w'$ is the walk obtained by concatenation of $w'$ to
the end of $w$ with the first and second literals of all entries in
$w$ and $w'$ preserving their roles in $w+w'$.

\begin{definition}
A path of a 2-CNF formula $F$ is a walk of $F$ all clauses of which
are pairwise distinct.
\end{definition}

Consider an example demonstrating the above notions. Let $w=(l_1
\vee l_2),(\neg l_2 \vee l_3),(\neg l_3 \vee l_4),(\neg l_4 \vee
\neg l_3),(l_3 \vee \neg l_2),(l_2 \vee l_5)$ be a walk of some
2-CNF formula presented so that the first literals of all entries
appear before the second literals. Then $l_1$ and $l_5$ are the
first and the last literals of $w$, respectively, and hence $w$ is a
walk from $l_1$ to $l_5$. The clause $(\neg l_2 \vee l_3)$ has an
interesting property that both its literals are first literals of
this clause with respect to $w$ (and therefore the second literals
as well). The second item of $w$ witnesses $\neg l_2$ being a first
literal of $(\neg l_2 \vee l_3)$ w.r.t. $w$ (and hence $l_3$ being a
second one), while the second item of $w$ from the end provides the
witness for $l_3$ being a first literal of $(\neg l_2 \vee l_3)$
w.r.t. $w$ (and hence $\neg l_2$ being a second one). The rest of
clauses do not possess this property. For example $l_1$ is the first
literal of $(l_1 \vee l_2)$ w.r.t. $w$ (as witnessed by the first
entry) but not the second one. Next, $reverse(w)=(l_5 \vee
l_2),(\neg l_2 \vee l_3),(\neg l_3 \vee \neg l_4),(l_4 \vee \neg
l_3),(l_3 \vee \neg l_2),(l_2 \vee l_1)$. Let $w_1$ be the prefix of
$w$ containing all the clauses except the last one. Then $w=w_1+(l_2
\vee l_5)$. Let $w_2$ be the prefix of $w$ containing the first 4
entries, $w_3$ be the suffix of $w$ containing the last 2 entries.
Then $w=w_2+w_3$. Finally, observe that $w$ is not a path due to the
repeated occurrence of clause $(\neg l_2 \vee l_3)$, while $w_2$ is
a path.

\subsection{2-ASAT and 2-ASLASAT problems.}
\label{probdef}
\begin{definition} \label{problems}
\begin{enumerate}
\item A Culprit Set (CS) of a 2-CNF formula $F$ is a subset $S$ of
$Clauses(F)$ such that $F \setminus S$ is satisfiable.
\item Let $(F,L,l)$ be a triple where $F$ is a 2-CNF formula, $L$ is a
non-contradictory set of literals such that $SWRT(F,L)$ is true and
$l$ s a literal such that $Var(l) \notin Var(L)$. A CS of $(F,L,l)$
is a subset $S$ of $Clauses(F)$ such that $SWRT(F \setminus S, L
\cup \{l\})$ is true.
\end{enumerate}
\end{definition}

Having defined a CS with respect to two different structures, we
define problems of finding a smallest CS (SCS) with respect to these
structures. In particular \emph{Almost 2-SAT problem} (2-ASAT
problem) is defined as follows: given a 2-CNF formula $F$, find an
SCS of $F$. The \emph{Annotated Almost 2-SAT problem with single
literal} (2-ASLASAT problem) is defined as follows: given the
triplet $(F,L,l)$ as in the last item of Definition \ref{problems},
find an SCS of $(F,L,l)$.

Now we introduce parameterized versions of the 2-ASAT and 2-ASLASAT
problems, where the parameter restricts the size of a CS. In
particular, the input of the \emph{parameterized 2-ASAT} problem is
$(F,k)$, where $F$ is a 2-CNF formula and $k$ is a non-negative
integer. The output is a CS of $F$ of size at most $k$, if one
exists. Otherwise, the output is 'NO'. The input of the
\emph{parameterized 2-ASLASAT} problem is $(F,L,l,k)$ where
$(F,L,l)$ is as specified in Definition \ref{problems}. The output
is a CS of $(F,L,l)$ of size at most $k$, if there is such one.
Otherwise, the output is 'NO'.

\section{2-ASLASAT problem: related theorems.}
\subsection{Basic Lemmas.}
\begin{lemma} \label{impl}
Let $F$ be a \textsc{2-cnf} formula and $w$ be a walk of $F$. Let
$l_x$ and $l_y$ be the first and the last literals of $w$,
respectively. Then $SWRT(F, \{\neg l_x, \neg l_y\})$ is false. In
particular, if $l_x=l_y$ then $SWRT(F, \neg l_x)$ is false.
\end{lemma}

{\bf Proof.} Since $w$ is a walk of $F$, $Var(l_x) \in Var(F)$ and
$Var(l_y) \in Var(F)$. Consequently for any satisfying assignment
$P$ of $F$ both $Var(l_x)$ and $Var(l_y)$ belong to $Var(P)$.
Therefore $SWRT(F,\{\neg l_x, \neg l_y\})$ may be true only if there
is a satisfying assignment of $F$ containing both $\neg l_x$ and
$\neg l_y$. We going to show that this is impossible by induction on
the length of $w$ This is clear if $|w|=1$ because in this case
$w=(l_x \vee l_y)$. Assume that $|w|>1$ and the statement is
satisfied for all shorter walks. Then $w=w'+(l_t \vee l_y)$, where
$w'$ is a walk of $w$ from $l_x$ to $\neg l_t$. By the induction
assumption $SWRT(F,\{\neg l_x,l_t\})$ is false and hence any
satisfying assignment of $F$ containing $\neg l_x$ contains $\neg
l_t$ and hence contains $l_y$. As we noted above in the proof, this
implies that $SWRT(F, \{\neg l_x, \neg l_y\})$ is false.
$\blacksquare$

\begin{lemma} \label{singlelit}
Let $F$ be a \textsc{2-cnf} formula and let $L$ be a set of literals
such that $SWRT(F,L)$ is true. Let $C=(l_1 \vee l_2)$ be a clause of
$F$ and let $w$ be a walk of $F$ from $\neg L$ containing $C$ and
assume that $l_1$ is a first literal of $C$ w.r.t. $w$. Then $l_1$
is \emph{not} a second literal of $C$ w.r.t. any walk from $\neg L$.
\end{lemma}

{\bf Proof.} Let $w'$ be a walk of $F$ from $\neg L$ which contains
$C$ so that $l_1$ is a second literal of $C$ w.r.t. $w'$. Then $w'$
has a prefix $w''$ whose last literal is $l_1$. Let $l'$ be the
first literal of $w'$ (and hence of $w''$). According to Lemma
\ref{impl}, $SWRT(F, \{\neg l_1,\neg l'\})$ is false. Therefore if
$l_1 \in \neg L$ then $SWRT(F,L)$ is false (because $\{\neg l_1,
\neg l'\} \subseteq L$) in contradiction to the conditions of the
lemma. Thus $l_1 \notin \neg L$ and hence $l_1$ is not the first
literal of $w$. Consequently, $w$ has a prefix $w^*$ whose last
literal is $\neg l_1$. Let $l^*$ be the first literal of $w$ (and
hence of $w^*$). Then $w^*+reverse(w'')$ is a walk from $l^*$ to
$l'$, both belong to $\neg L$. According to Lemma \ref{impl},
$SWRT(F,\{\neg l^*,\neg l'\})$ is false and hence $SWRT(F,L)$ is
false in contradiction to the conditions of the lemma. It follows
that the walk $w'$ does not exist and the present lemma is correct.
$\blacksquare$

\begin{lemma} \label{walkpath}
Let $F$ be a \textsc{2-cnf} formula, let $L$ be a set of literals
such that $SWRT(F,L)$ is true, and let $w$ be a walk from $\neg L$.
Then $F$ has a path $p$ with the same first and last literals as $w$
and the set of clauses of $p$ is a subset of the set of clauses of
$w$.
\end{lemma}
 {\bf Proof.} The proof is by induction on the length of $w$. The
statement is clear if $|w|=1$ because $w$ itself is the desired
path. Assume that $|w|>1$ and the lemma holds for all shorter paths
from $\neg L$. If all clauses of $w$ are distinct then $w$ is the
desired path. Otherwise, let $w=(C_1, \dots, C_q)$ and assume that
$C_i=C_j$ where $1 \leq i<j \leq q$. By Lemma \ref{singlelit}, $C_i$
and $C_j$ have the same first (and, of course, the second) literal.
If $i=1$, let $w'$ be the suffix of $w$ starting at $C_j$.
Otherwise, if $C_j=q$, let $w'$ be the prefix of $w$ ending at
$C_i$. If none of the above happens then $w'=(C_1, \dots,
C_{i},C_{j+1},C_q)$. In all the cases, $w'$ is a walk of $F$ with
the same first and last literals as $w$ such that $|w'|<|w|$ and the
set of clauses of $w'$ is a subset of the set of clauses of $w$. The
desired path is extracted from $w'$ by the induction assumption.
$\blacksquare$

\subsection{A non-empty SCS of $(F,L,l)$: necessary and sufficient
condition}

\begin{theorem}\label{insatisf}
Let $(F,L,l)$ be an instance of the 2-ASLASAT problem. Then
$SWRT(F,L \cup \{l\})$ is false if and only if $F$ has a walk from
$\neg l$ to $\neg l$ or a walk from $\neg L$ to $\neg l$.
\end{theorem}

{\bf Proof.} Assume that $F$ has a walk from $\neg l$ to $\neg l$ or
from $\neg l'$ to $\neg l$ such that $l' \in L$. Then, according to
Lemma \ref{impl}, $SWRT(F,l)$ is false or $SWRT(F, \{l',l\})$ is
false, respectively. Clearly in both cases $SWRT(F,L \cup \{l\})$ is
false as $L \cup \{l\}$ is, by definition, a superset of both
$\{l\}$ and $\{l',l\}$.

Assume now that $SWRT(F,L \cup \{l\})$ is false. Let $I$ be a set of
literals including $l$ and all literals $l'$ such that $F$ has a
walk from $\neg l$ to $l'$. Let $S$ be the set of all clauses of $F$
satisfied by $I$.

Assume that $I$ is non-contradictory and does not intersect with
$\neg L$. Let $P$ be a satisfying assignment of $F$ which does not
intersect with $\neg L$ (such an assignment exists according to
definition of the \textsc{2-aslasat} problem). Let $P'$ be the
subset of $P$ such that $Var(P')=Var(F) \setminus Var(I)$. Observe
that $P' \cup I$ is non-contradictory. Indeed, $P'$ is
non-contradictory as being a subset of a satisfying assignment $P$
of $F$, $I$ is non-contradictory by assumption, and due to the
disjointness of $Var(I)$ and $Var(P')$, there is no literal $l' \in
I$ and $\neg l' \in P'$. Next, note that every clause $C$ of $F$ is
satisfied by $P' \cup I$. Indeed, if $C \in S$ then $C$ is satisfied
by $I$, by definition of $I$. Otherwise, assume first that $Var(C)
\cap Var(I) \neq \emptyset$. Then $C=(\neg l' \vee l'')$, where $l'
\in I$. Then either $l'=l$ or $F$ has a walk $w$ from $\neg l$ to
$l'$. Consequently, either $(\neg l' \vee l'')$ or $w+(\neg l' \vee
l'')$ is a walk from $\neg l$ to $l''$ witnessing that $l'' \in I$
and hence $C \in S$, a contradiction. It remains to conclude that
$Var(C) \cap Var(I)=\emptyset$, i.e. that $Var(C) \subseteq
Var(P')$. If $P'$ contains contradictions of both literals of $C$
then $P \setminus P'$ contains at least one literal of $C$ implying
that $P$ contains a literal and its negation in contradiction to the
definition of $P$. Consequently, $C$ is satisfied by $P'$. Taking
into account that $Var(P' \cup I)=Var(F)$, $P' \cup I$ is a
satisfying assignment of $F$. Observe that $P' \cup I$ does not
intersect with $\neg(L \cup l)$. Indeed, both $I$ and $P'$ do not
intersect with $\neg L$, the former by assumption the latter by
definition. Next, $l \in I$ and $P' \cup I$ is non-contradictory,
hence $\neg l \notin P' \cup I$. Thus $P' \cup I$ witnesses that
$SWRT(F, L \cup \{l\})$ is true in contradiction to our assumption.
Thus our assumption regarding $I$ made in the beginning of the
present paragraph is incorrect.

It follows from the previous paragraph that either $I$ contains a
literal and its negation or $I$ intersects with $\neg L$. In the
former case if $\neg l \in I$ then by definition of $I$ there is a
walk from $\neg l$ to $\neg l$. Otherwise $I$ contains $l'$ and
$\neg l'$ such that $Var(l') \neq Var(l)$. Let $w_1$ be the walk
from $\neg l$ to $l'$ and let $w_2$ be the walk from $\neg l$ to
$\neg l'$ (both walks exist according tot he definition of $I$).
Clearly $w_1+reverse(w_2)$ is a walk from $\neg l$ to $\neg l$. In
the latter case, $F$ has a walk $w$ from $\neg l$ to $\neg l'$ such
that $l' \in L$. Clearly $reverse(w)$ is a walk from $\neg L$ to
$\neg l$. Thus we have shown that if $SWRT(F,L \cup \{l\})$ is false
then $F$ has a walk from $\neg l$ to $\neg l$ or a walk from $\neg
L$ to $\neg l$, which completes the proof of the theorem.
$\blacksquare$

\subsection{Smallest Separators}

\begin{definition} \label{gensep}
A set $SC$ of clauses of a 2-CNF formula $F$ is a separator with
respect to a set of literals $L$ and literal $l_y$ if $F \setminus
SC$ does not have a path from $L$ to $l_y$.
\end{definition}

We denote by $SepSize(F,L,l_y)$ the size of a smallest separator of
$F$ w.r.t. $L$ and $l_y$ and by ${\bf OptSep}(F,L,l_y)$ the set of
all smallest separators of $F$ w.r.t. $L$ and $l_y$. Thus for any $S
\in {\bf OptSep}(F,L,l_y)$, $|S|=SepSize(F,L,l_y)$.

Given the above definition, we derive an easy corollary from Lemma
\ref{impl}.

\begin{corollary} \label{lowerbound}
Let $(F,L,l)$ be an instance of the 2-ASLASAT problem. Then the size
of an SCS of this instance is greater than or equal to $SepSize(F,
\neg L,\neg l)$.
\end{corollary}

{\bf Proof.} Assume by contradiction that $S$ is a CS of $(F,L,l)$
such that $|S|<SepSize(F, \neg L, \neg l)$. Then $F \setminus S$ has
at least one path $p$ from a literal $\neg l'$ ($l' \in L$) to $\neg
l$. According to Lemma \ref{impl}, $F \setminus S$ is not
satisfiable w.r.t. $\{l',l\}$ and hence it is not satisfiable with
respect to $L \cup \{l\}$ which is a superset of $\{l',l\}$. That
is, $S$ is \emph{not} a CS of $(F,L,l)$, a contradiction.
$\blacksquare$
% A short proof of Theorem \ref{genmenger} which do not require Theorem \ref{mengerlike}

%This should be postponed to the preliminaries
Let $D=(V,A)$ be the \emph{implication graph} on $F$ which is a
digraph whose set $V(D)$ of nodes corresponds to the set of literals
of the variables of $F$ and $(l_1,l_2)$ is an arc in its set $A(D)$
of arcs if and only if $(\neg l_1 \vee l_2) \in Clauses(F)$. We say
that arc $(l_1,l_2)$ \emph{represents} the clause $(\neg l_1 \vee
l_2)$. Note that each arc represents exactly one clause while a
clause including two distinct literals is represented by two
different arcs. In particular, if $\neg l_1 \neq l_2$, the other arc
which represents $(\neg l_1 \vee l_2)$ is $(\neg l_2,\neg l_1)$. In
the context of $D$ we denote by $L$ and $\neg L$ the set of nodes
corresponding to the literals of $L$ and $\neg L$, respectively. We
adopt the definition of a walk and a path of a digraph given in
\cite{Gutinbook}. Taking into account that all the walks of $D$
considered in this paper are non-empty we represent them as the
sequences of arcs instead of alternative sequences of arcs and
nodes. In other words, if $w=(x_1,e_1, \dots, x_q,e_q,x_{q+1})$ is a
walk of $D$, we represent it as $(e_1, \dots, e_q)$. The \emph{arc
separator} of $D$ w.r.t. a set of literals $L$ and a literal $l$ is
a set of arcs such that the graph resulting from their removal has
no path from $L$ to $l$. Similarly to the case with \textsc{2-cnf}
formulas, we denote by $ArcSepSize(D,L,l)$ the size of the smallest
arc separator of $D$ w.r.t. $L$ and $l$.

\begin{theorem} \label{genmenger}
Let $F$ be a \textsc{2-cnf} formula, let $L$ be a set of literals
such that $SWRT(F, \neg L)$ is true. Let $l_y$ be a literal such
that $Var(l_y) \notin Var(L)$. Then the following statements hold.
\begin{enumerate}
\item The largest number $MaxPaths(F,L,l_y)$ of clause-disjoint paths from $L$ to
$l_y$ in $F$ equals the largest number $MaxPaths(D, \neg L,l_y)$ of
arc-disjoint paths from $\neg L$ to $l_y$ in $D$.
\item $SepSize(F,L,l_y)=ArcSepSize(D, \neg L,l_y)$
\item $MaxPaths(F,L,l_y)=SepSize(F,L,l_y)$.
\end{enumerate}
\end{theorem}

Note that generally (if there is no requirement that $SWRT(F, \neg
L)$ is true) $SepSize(F,L,l_y)$ may differ from $ArcSepSize(D, \neg
L,l_y)$. The reason is that a separator of $D$ may correspond to a
smaller separator of $F$ due to the fact that some arcs may
represent the same clause. As we will see in the proof, the
requirement that $SWRT(F, \neg L)$ is true rules out this
possibility.

{\bf Proof of Theorem \ref{genmenger}.} We may safely assume that
$Var(L) \subseteq Var(F)$ because literals whose variables do not
belong to $Var(F)$ cannot be starting points of paths in $F$. Also
since $l_y \notin \neg L$ any walk from $\neg L$ to $l_y$ in $D$ is
non-empty. We use this fact implicitly in the proof without
referring to it.

Let $w=(C_1, \dots, C_q)$ be a walk from $l'$ to $l''$ in $F$. Let
$w(D)=(a_1, \dots, a_q)$ be the sequence of arcs of $D$ constructed
as follows. For each $C_i=(l_1 \vee l_2)$ (we assume that $l_1$ is
the first literal of $C_i$), $a_i=(\neg l_1,l_2)$. Then $\neg l'$ is
the tail of $a_1$ and $l''$ is the head of $a_q$. Also, by
definition of $w$, for any two arcs $a_i$ and $a_{i+1}$, the head of
$a_i$ is the same as the tail of $a_{i+1}$. It follows that $w(D)$
is a walk from $\neg l'$ to $l''$ in $D$ such that each $a_i$
represents $C_i$. Now, let ${\bf P}=\{p_1, \dots, p_t\}$ be a set of
clause-disjoint paths from $L$ to $l_y$ in $F$. Then $\{p_1(D),
\dots, p_q(D)\}$ is a set of walks from $\neg L$ to $l_y$ in $D$
which are arc-disjoint. Indeed, if an arc $a$ belongs to both
$p_i(D)$ and $p_j(D)$ (where $i \neq j$) then, due to the
disjointness of $p_i$ and $p_j$, this arc $a$ represents two
different clauses which is impossible by definition. Since every
$p_i(D)$ includes a path $p'_i(D)$ with the same first and last
nodes and the set of arcs being a subset of the set of arcs of
$p_i(D)$ (see \cite{Gutinbook}, Proposition 4.1.), we can specify
$t$ arc-disjoint paths $\{p'_1(D), \dots p'_t(D)\}$ from $\neg L$ to
$l_y$, which shows that $MaxPaths(D, \neg L,l_y) \geq
MaxPaths(F,L,l_y)$.

Conversely, let $p=(a_1, \dots, a_q)$ be a path from $\neg l'$ to
$l''$ in $D$. Let $p(F)$ be the sequence $(C_1, \dots, C_q)$ of
clauses defined as follows. For each $a_i=(\neg l_1,l_2)$, $C_i=(l_1
\vee l_2)$, $l_1$ and $l_2$ are specified as the first and the
second literals of $C_i$, respectively. Then $l'$ is the first
literal of $C_1$, $l''$ is the last literal of $C_q$ and for each
consecutive pair $C_i$ and $C_{i+1}$ the second literal of $C_i$ is
the negation of the first literal of $C_{i+1}$. In other words,
$p(F)$ is a walk from $l'$ to $l''$ in $F$ where each $C_i$ is
represented by $a_i$. Now, let ${\bf P}=\{p_1, \dots, p_t\}$ be a
set of arc-disjoint paths from $\neg L$ to $l_y$ in $D$. Then
$\{p_1(F), \dots p_t(F)\}$ is a set of walks from $L$ to $l_y$ in
$F$. Observe that these walks are clause-disjoint. Indeed, if a
clause $C=(l_1 \vee l_2)$ belongs to both $p_i(F)$ and $p_j(F)$
(where $i \neq j$) then $(l_1 \vee l_2)$ is represented by arc, say,
$(\neg l_1,l_2)$ in $p_i$ and by arc $(\neg l_2,l_1)$ in $p_j$. By
construction of $p_i(F)$ and $p_j(F)$, $l_1$ is the first literal of
$C$ w.r.t. $p_i(F)$ and the second literal of $C$ w.r.t. $p_j(F)$
which contradicts Lemma \ref{singlelit}. That is the walks of
$\{p_1(F), \dots, p_t(F)\}$ are clause-disjoint. Also, by Lemma
\ref{walkpath}, for each $p_i(F)$, there is a path $p'_i(F)$ of $F$
with the same first and last literals as $p_i(F)$ and whose set of
clauses is a subset of the set of clauses of $p_i(F)$. Clearly the
paths $\{p'_1(F), \dots, p'_t(F)\}$ are clause disjoint. Thus
$MaxPaths(D, \neg L,l_y) \leq MaxPaths(F,L,l_y)$. Combining this
statement with the statement proven in the previous paragraph, we
conclude that $MaxPaths(D, \neg L,l_y) = MaxPaths(F,L,l_y)$.

Let $S \in {\bf OptSep}(F,L,l_y)$. For each $C \in S$, let $p_C$ be
a path of $F$ from $L$ to $l_y$ including $C$ (such a path
necessarily exists due to the minimality of $S$). Let $a(C)$ be an
arc of $p_C(D)$ which represents $C$. Let $S(D)$ be the set of all
$a(C)$. We are going to show that $S(D)$ separates $\neg L$ from
$l_y$ in $D$. Assume that this is not so and let $p^*$ be a path
from $\neg L$ to $l_y$ in $D \setminus S(D)$. Then, according to
Lemma \ref{walkpath}, $p^*(F)$ necessarily includes a path from $L$
to $l_y$ and hence $p^*(F)$ contains at least one clause $C=(l_1
\vee l_2)$ of $S$. Let $a^*$ be an arc of $p^*$ which represents
$C$. By definition of of $p^*$, $a^* \neq a(C)$ and hence $a(C)$ is,
say $(\neg l_1,l_2)$ and $a^*$ is $(\neg l_2,l_1)$. By definition of
$p_C(D)$ and $p^*(F)$, $l_1$ is the first literal of $C$ w.r.t.
$p_C$ and the second one w.r.t. $p^*(F)$ which contradicts Lemma
\ref{singlelit}. This shows that $S(D)$ separates $\neg L$ from
$l_y$ in $D$ and, consequently, taking into account that
$|S(D)|=|S|$, $ArcSepSize(D, \neg L,l_y) \leq SepSize(F,L,l_y)$.

Let $S$ be a smallest arc separator of $D$ w.r.t. $\neg L$ and
$l_y$. For each $a \in S$, let $p_a$ be a path of $D$ from $\neg L$
to $l_y$ which includes $a$. Let $C(a)$ be a clause of $p_a(F)$
which is represented by $a$. Denote the set of all $C(a)$ by $S(F)$.
Then we can show that $S(F)$ is a separator w.r.t. $L$ and $l_y$ in
$F$. In particular, let $p^*$ be a path from $L$ to $l_y$ in $F
\setminus S(F)$. Then $p^*(D)$ necessarily includes an arc $a \in
S$. Let $C^*$ be a clause of $p^*$ represented by $a$. Since $C^*
\neq C(a)$, the arc $a$ represents two different clauses in
contradiction to the definition of $D$. Consequently, taking into
account that $|S(F)| \leq |S|$, $ArcSepSize(D, \neg L,l_y) \geq
SepSize(F,L,l_y)$. Considering the previous paragraph we conclude
that $ArcSepSize(D, \neg L,l_y) = SepSize(F,L,l_y)$.

Let ${\bf PF}$ be a largest set of clause-disjoint paths from $L$ to
$l_y$ in $F$ and let ${\bf PD}$ be a largest set of arc-disjoint
paths from $\neg L$ to $l_y$ in $D$. It follows from the above proof
that in order to show that $|{\bf PF}|=SepSize(F,L,l_y)$, it is
sufficient to show that $|{\bf PD}|=ArcSepSize(D,\neg L,l_y)$.
Taking into account that by our assumption $l_y \notin \neg L$, the
latter can be easily derived by contracting the vertices of $\neg L$
into one vertex and applying the arc version of Menger's Theorem for
directed graphs \cite{Gutinbook}. $\blacksquare$

%%%%%%%%%%%%%%%%%%%%%%%%%%%%%%%%%%%%%%%%%%%%%%%%%%%%%%%%%%%%%%%%%%%%%%%%%%%%%%%%%%%%%%

\subsection{Neutral Literals}

\begin{definition}
Let $(F,L,l)$ be an instance of the \textsc{2-aslasat} problem. A
literal $l^*$ is a \emph{neutral literal} of $(F,L,l)$ if $(F, L
\cup \{l^*\},l)$ is a valid instance of \textsc{2-aslasat} problem
and $SepSize(F,\neg L,\neg l)=SepSize(F,\neg (L \cup \{l^*\}),\neg
l)$.
\end{definition}

The following theorem has a crucial role in the design of the
algorithm provided in the next section.

\begin{theorem} \label{pseudo}
Let $(F,L,l)$ be an instance of the 2-ASALSAT problem and let $l^*$
be a neutral literal of $(F,L,l)$. Then there is a CS of $(F,L \cup
\{l^*\},l)$ of size smaller than or equal to the size of an SCS of
$(F,L,l)$.
\end{theorem}

Before we prove Theorem \ref{pseudo}, we extend our terminology.

\begin{definition} \label{reach}
Let $(F,L,l)$ be an instance of the 2-ASLASAT problem. A clause
$C=(l_1 \vee l_2)$ of $F$ is \emph{reachable} from $\neg L$ if there
is a walk $w$ from $\neg L$ including $C$. Assume that $l_1$ is a
first literal of $C$ w.r.t. $w$. Then $l_1$ is called \emph{the
main} literal of $C$ w.r.t. $(F,L,l)$.
\end{definition}

Given Definition \ref{reach}, Lemma \ref{singlelit} immediately
implies the following corollary.

\begin{corollary} \label{mainlit}
Let $(F,L,l)$ be an instance of the 2-ASLASAT problem and let
$C=(l_1 \vee l_2)$ be a clause reachable from $\neg L$. Assume that
$l_1$ is the main literal of $C$ w.r.t. $(F,L,l)$. Then $l_1$ is not
a second literal of $C$ w.r.t. any walk $w'$ starting from $\neg L$
and including $C$.
\end{corollary}

Now we are ready to prove Theorem \ref{pseudo}.

{\bf Proof of Theorem \ref{pseudo}.} Let $SP \in {\bf OptSep}(F,
\neg( L \cup \{l^*\}),\neg l)$. Since $\neg L$ is a subset of $\neg(
L \cup \{l^*\})$, $SP$ is a separator w.r.t. $\neg L$ and $\neg l$
in $F$.  Moreover, since $l^*$ is a neutral literal of $(F,L,l)$,
$SP \in {\bf OptSet}(F, \neg L, \neg l)$.

In the \textsc{2-cnf} $F \setminus SP$, let $R$ be the set of
clauses reachable from $\neg L$ and let $NR$ be the rest of the
clauses of $F \setminus SP$. Observe that the sets $R,NR,SP$ are a
partition of the set of clauses of $F$.

Let $X$ be a SCS of $(F,L,l)$. Denote $X \cap R$, $X \cap SP$, $X
\cap NR$ by $XR$, $XSP$, $XNR$ respectively. Observe that the sets
$XR,XSP,XNR$ are a partition of $X$.

Let $Y$ be the subset of $SP \setminus XSP$ including all clauses
$C=(l_1 \vee l_2)$ (we assume that $l_1$ is the main literal of $C$)
such that there is a walk $w$ from $l_1$ to $\neg l$ with $C$ being
the first clause of $w$ and all clauses of $w$ following $C$ (if
any) belong to $NR \setminus XNR$. We call this walk $w$ \emph{a
witness walk} of $C$. By definition, $SP \setminus XP=SP \setminus
X$ and $NR \setminus XNR=NR \setminus X$, hence the clauses of $w$
do not intersect with $X$.

\begin{cclaim} \label{sizey}
$|Y| \leq |XR|$.
\end{cclaim}

{\bf Proof.} By definition of the \textsc{2-aslasat} problem,
$SWRT(F,L)$ is true. Therefore, according to Theorem
\ref{genmenger}, there is a set ${\bf P}$ of $|SP|$ clause-disjoint
paths from $\neg L$ to $\neg l$. Clearly each $C \in SP$
participates in exactly one path of ${\bf P}$ and each $p \in {\bf
P}$ includes exactly one clause of $SP$. In other words, we can make
one-to-one correspondence between paths of ${\bf P}$ and the clauses
of $SP$ they include. Let ${\bf PY}$ be the subset of ${\bf P}$
consisting of the paths corresponding to the clauses of $Y$. We are
going to show that for each $p \in {\bf PY}$ the clause of $SP$
corresponding to $p$ is preceded in $p$ by a clause of $XR$.

Assume by contradiction that this is not true for some $p \in {\bf
PY}$ and let $C=(l_1 \vee l_2)$ be the clause of $SP$ corresponding
to $p$ with $l_1$ being the main literal of $C$ w.r.t. $(F,L,l)$. By
our assumption, $C$ is the only clause of $SP$ participating in $p$,
hence all the clauses of $p$ \emph{preceding} $C$ belong to $R$.
Consequently, the only possibility of those preceding clauses to
intersect with $X$ is intersection with $XR$. Since this possibility
is ruled out according to our assumption, we conclude that no clause
of $p$ preceding $C$ belongs to $X$. Next, according to Corollary
\ref{mainlit}, $l_1$ is the first literal of $C$ w.r.t $p$, hence
the suffix of $p$ starting at $C$ can be replaced by the witness
walk of $C$ and as a result of this replacement, a walk $w'$ from
$\neg L$ to $\neg l$ is obtained. Taking into account that the
witness walk of $C$ does not intersect with $X$, we get that $w'$
does not intersect with $X$. By Theorem \ref{insatisf}, $SWRT(F
\setminus X, L \cup \{l\})$ is false in contradiction to being $X$ a
CS of $(F,L,l)$. This contradiction shows that our initial
assumption fails and $C$ is preceded in $p$ by a clause of $XR$.

In other words, each path of ${\bf PY}$ intersects with a clause of
$XR$. Since the paths of ${\bf PY}$ are clause-disjoint, $|XR| \geq
|{\bf PY}|=|Y|$, as required. $\square$

Consider the set $X^*=Y \cup XSP \cup XNR$. Observe that
$|X^*|=|Y|+|XSP|+|XNR| \leq |XR|+|XSP|+|XNR|=|X|$, the first
equality follows from the mutual disjointness of $Y$, $XSP$ and
$XNR$ by their definition, the inequality follows from Claim
\ref{sizey}, the last equality was justified in the paragraph where
the sets $XP$, $XSP$, $XNR$, and $X$ have been defined. We are going
to show that $X^*$ is a CS of $(F, L \cup\{l^*\},l)$ which will
complete the proof of the present theorem.

\begin{cclaim} \label{nowalk1}
$F \setminus X^*$ has no walk from $\neg (L \cap \{l^*\})$ to $\neg
l$.
\end{cclaim}

{\bf Proof.} Assume by contradiction that $w$ is a walk from $\neg
(L \cap \{l^*\})$ to $\neg l$ in $F \setminus X^*$. Taking into
account that $SWRT(F \setminus X^*, L \cup \{l^*\})$ is true
(because we know that $SWRT(F, L \cup \{l^*\})$ is true), and
applying Lemma \ref{walkpath}, we get that $F \setminus X^*$ has a
path $p$ from $\neg (L \cap \{l^*\})$ to $\neg l$. As $p$ is a path
in $F$, it includes at least one clause of $SP$ (recall that $SP$ is
a separator w.r.t. $\neg (L \cap \{l^*\})$ and $\neg l$ in $F$). Let
$C=(l_1 \vee l_2)$ be the last clause of $SP$ as we traverse $p$
from $\neg (L \cap \{l^*\})$ to $\neg l$ and assume w.l.o.g. that
$l_1$ is the main literal of $C$ w.r.t. $(F \setminus X^*,L \cup
\{l^*\},l)$ (and hence of $(F,L \cup \{l^*\},l)$). Let $p^*$ be the
suffix of $p$ starting at $C$.

According to Corollary \ref{mainlit}, $l_1$ is the first literal of
$p^*$. In the next paragraph we will show that no clause of $R$
follows $C$ is $p^*$. Combining this statement with the observation
that the clauses of $F \setminus X^*$ can be partitioned into $R$,
$SP \setminus XSP$ and $NR \setminus XNR$ (the rest of clauses
belong to $X^*$) we conclude that $p^*$ is a walk witnessing that $C
\in Y$. But this is a contradiction because by definition $Y
\subseteq X^*$. This contradiction will complete the proof of the
present claim.

Assume by contradiction that $C$ is followed in $p^*$ by a clause
$C'=(l'_1 \vee l'_2)$ of $R$ (we assume w.l.o.g. that $l'_1$ is the
main literal of $C'$ w.r.t. $(F \setminus X^*,L \cup \{l^*\},l)$).
Let $p'$ be a suffix of $p^*$ starting at $C'$. It follows from
Corollary \ref{mainlit} that the first literal of $p'$ is $l'_1$. By
definition of $R$ and taking into account that $R \cap
X^*=\emptyset$, $F \setminus X^*$ has a walk $w_1$ from $\neg L$
whose last clause is $C'$ and all clauses of which belong to $R$. By
Corollary \ref{mainlit}, the last literal of $w_1$ is $l'_2$.
Therefore we can replace $C'$ by $w_1$ in $p'$. As a result we get a
walk $w_2$ from $\neg L$ to $\neg l$ in $F \setminus X^*$. By Lemma
\ref{walkpath}, there is a path $p_2$ from $\neg L$ to $\neg l$
whose set of clauses is a subset of the set of clauses of $w_2$. As
$p_2$ is also a path of $F$, it includes a clause of $SP$. However,
$w_1$ does not include any clause of $SP$ by definition. Therefore,
$p'$ includes a clause of $SP$. Consequently, $p^*$ includes a
clause of $SP$ following $C$ in contradiction to the selection of
$C$. This contradiction shows that clause $C'$ does not exist, which
completes the proof of the present claim as noted in the previous
paragraph. $\square$

\begin{cclaim} \label{nowalk2}
$F \setminus X^*$ has no walk from $\neg l$ to $\neg l$.
\end{cclaim}

{\bf Proof.} Assume by contradiction that $F \setminus X^*$ has a
walk $w$ from $\neg l$ to $\neg l$. By definition of $X$ and Theorem
\ref{insatisf}, $w$ contains at least one clause of $X$. Since $XSP$
and $XNR$ are subsets of $X^*$, $w$ contains a clause $C'=(l'_1 \vee
l'_2)$ of $XR$. Assume w.l.o.g. that $l'_1$ is the main literal of
$C'$ w.r.t. $(F,L,l)$. If $l'_1$ is a first literal of $C'$ w.r.t.
$w$ then let $w^*$ be a suffix of $w$ whose first clause is $C'$ and
first literal is $l'_1$. Otherwise, let $w^*$ be a suffix of
$reverse(w)$ having the same properties. In any case, $w^*$ is a
walk from $l'_1$ to $\neg l$ in $F \setminus X^*$ whose first clause
is $C'$. Arguing as in the last paragraph of proof of Claim
\ref{nowalk1}, we see that $F \setminus X^*$ has a walk $w_1$ from
$\neg L$ to $l'_2$ whose last clause is $C'$. Therefore we can
replace $C'$ by $w_1$ in $w^*$ and get a walk $w_2$ from $\neg L$ to
$\neg l$ in $F \setminus X^*$ in contradiction to Claim
\ref{nowalk1}. This contradiction shows that our initial assumption
regarding the existence of $w$ is incorrect and hence completes the
proof of the present claim. $\square$

It follows from Combination of Theorem \ref{insatisf}, Claim
\ref{nowalk1}, and Claim \ref{nowalk2} that $X^*$ is a CS of $(F,L
\cup \{l^*\},l)$, which completes the proof of the present theorem.
$\blacksquare$

\section{Algorithm for the parameterized 2-ASLASAT problem and its analysis}
\subsection{The algorithm}
{\small
$\textsc{FindCS}(F,L,l,k)$\\
{\bf Input:} An instance $(F,L,l,k)$ of the parameterized \textsc{2-aslasat} problem.\\ %may be, I have to write $(F,L,l)$ is an instance of the 2-aslasat
{\bf Output:} A CS of $(F,L,l)$ of size at most $k$ if one exists.
Otherwise 'NO' is returned.
\begin{enumerate}
\item {\bf if} $SWRT(F,L \cup \{l\})$ is true {\bf then}
return $\emptyset$ %1
\item {\bf if} $k=0$ {\bf then} Return 'NO' %2
\item {\bf if} $k \geq |Clauses(F)|$ {\bf then} return $Clauses(F)$ %3
\item {\bf if} $SepSize(F, \neg L, \neg l)>k$ {\bf then} return 'NO'
\footnote{The correctness of this step follows from Corollary
\ref{lowerbound}}
\item {\bf if} $F$ has a walk from $\neg L$ to $\neg l$ {\bf then}\\
~~~~~Let $C=(l_1 \vee l_2)$ be a clause such that $l_1 \in \neg L$
and $Var(l_2) \notin Var(L)$ %5
\item {\bf else} Let $C=(l_1 \vee l_2)$ be a clause which belongs to
a walk of $F$ from $\neg l$ to $\neg l$ and $SWRT(F,\{l_1,l_2\})$ is
true \footnote{Doing the analysis, we will prove that on Steps 5 and
6 $F$ has at least one clause with the required property}
\item {\bf if} Both $l_1$ and $l_2$ belong to $\neg (L \cup \{l\})$
{\bf then}  %7
\begin{itemize}
\item [7.1] $S \leftarrow \textsc{FindCS}(F \setminus C,L,l,k-1)$
\item [7.2] {\bf if} $S$ is not 'NO' {\bf then} Return $S \cup
\{C\}$
\item [7.3] Return 'NO'
\end{itemize}
\item {\bf if} Both $l_1$ and $l_2$ do not belong to $\neg (L \cup
\{l\})$ {\bf then}
\begin{itemize}
\item [8.1] $S_1 \leftarrow \textsc{FindCS}(F,L \cup \{l_1\},l,k)$
\item [8.2] {\bf if} $S_1$ is not 'NO' {\bf then} Return $S_1$
\item [8.3] $S_2 \leftarrow \textsc{FindCS}(F,L \cup \{l_2\},l,k)$
\item [8.4] {\bf if} $S_2$ is not 'NO' {\bf then} Return $S_2$
\item [8.5] $S_3 \leftarrow \textsc{FindCS}(F \setminus C,L,l,k-1)$
\item [8.6] {\bf if} $S_3$ is not 'NO' {\bf then} Return $S_3 \cup \{C\}$
\item [8.7] Return 'NO'
\end{itemize}
(In the rest of the algorithm we consider the cases where exactly
one literal of $C$ belongs to $\neg (L \cup \{l\})$. W.l.o.g. we
assume that this literal is $l_1$)
\item {\bf if} $l_2$ is not neutral in $(F,L,l)$
\begin{itemize}
\item [9.1] $S_2 \leftarrow \textsc{FindCS}(F,L \cup \{l_2\},l,k)$
\item [9.2] {\bf if} $S_2$ is not 'NO' {\bf then} Return $S_2$
\item [9.3] $S_3 \leftarrow \textsc{FindCS}(F \setminus C,L,l,k-1)$
\item [9.4] {\bf if} $S_3$ is not 'NO' {\bf then} Return $S_3 \cup \{C\}$
\item [9.5] Return 'NO'
\end{itemize}
\item Return $\textsc{FindCS}(F,L \cup \{l_2\},l,k)$
\end{enumerate}
}

\subsection{Additional Terminology and Auxiliary Lemmas}

In order to analyze the above algorithm, we extend our terminology.
Let us call a quadruple $(F,L,l,k)$ a \emph{valid input} if
$(F,L,l,k)$ is a valid instance of the parameterized 2-ASLASAT
problem (as specified in Section 2.3.).

Now we introduce the notion of the \emph{search tree} $ST(F,L,l,k)$
produced by $\textsc{FindCS}(F,L,l,k)$. The root of the tree is
identified with $(F,L,l,k)$. If $\textsc{FindCS}(F,L,l,k)$ does not
apply itself recursively then $(F,L,l,k)$ is the only node of the
tree. Otherwise the children of $(F,L,l,k)$ correspond to the inputs
of the calls applied \emph{within} the call
$\textsc{FindCS}(F,L,l,k)$. For example, if
$\textsc{FindCS}(F,L,l,k)$ performs Step 9 then the children of
$(F,L,l,k)$ are $(F,L \cup \{l_2\},l,k)$ and $(F \setminus C,
L,l,k-1)$. For each child $(F',L',l',k')$ of $(F,L,l,k)$, the
subtree of $ST(F,L,l,k)$ rooted by $(F',L',l',k')$ is
$ST(F',L',l',k')$. It is clear from the description of
$\textsc{FindCS}$ that the third item of a valid input is not
changed for its children hence in the rest of the section when we
denote a child or descendant of $(F,L,l,k)$ we will leave the third
item unchanged, e.g. $(F_1,L_1,l,k_1)$.

\begin{lemma} \label{exclause}
Let $(F,L,l,k)$ be a valid input. The $Solve2ASLASAT(F,L,l,k)$
succeeds to select a clause on Steps 5 and 6.
\end{lemma}

{\bf Proof.} Assume that $F$ has a walk from $\neg L$ to $\neg l$
and let $w$ be the shortest possible such walk. Let $l_1$ be the
first literal of $w$ and let $C=(l_1 \vee l_2)$ be the first clause
of $F$. By definition $l_1 \in \neg L$. We claim that $Var(l_2)
\notin Var(L)$. Indeed, assume that this is not true. If $l_2 \in
\neg L$ then $SWRT(F, \{\neg l_1,\neg l_2\})$ is false and hence
$SWRT(F,L)$ is false as $L$ is a superset of $\{\neg l_1,\neg
l_2\}$. But this contradicts the definition of  the
\textsc{2-aslasat} problem. Assume now that $l_2 \in L$. By
definition of the \textsc{2-aslasat} problem, $Var(l) \notin
Var(L)$, hence $C$ is not the last clause of $w$. Consequently the
first literal of the second clause of $w$ belongs to $\neg L$ . Thus
if we remove the first clause from $w$ we obtain a shorter walk from
$\neg L$ to $\neg l$ in contradiction to the definition of $w$. It
follows that our claim is true and the required clause $C$ can be
selected if the condition of Step 5 is satisfied.

Consider now the case where the condition of Step 5 is not
satisfied. Note that $SWRT(F, L \cup \{l\})$ is false because
otherwise the algorithm would have finished at Step 1. Consequently
by Theorem \ref{insatisf}, $F$ has a walk from $\neg l$ to $\neg l$.
We claim that any such walk $w$ contains a clause $C=(l_1 \vee l_2)$
such that $SWRT(F, \{l_1,l_2\})$ is true. Let $P$ be a satisfying
assignment of $F$ (which exists by definition of the
\textsc{2-aslasat} problem). Let $F'$ be the \textsc{2-cnf} formula
created by the clauses of $w$ and let $P'$ be the subset of $P$ such
that $Var(P')=Var(F')$. By Lemma \ref{impl}, $SWRT(F',l)$ is false
and hence, taking into account that $Var(l) \in Var(F')$, $\neg l
\in P'$. Consequently $l \in \neg P'$. Therefore $\neg P'$ is not a
satisfying assignment of $F'$ i.e. $\neg P'$ does not satisfy at
least one clause of $F'$. Taking into account that $Var(\neg
P')=Var(F')$, it contains negations of both literals of at least one
clause $C$ of $F'$. Therefore $P'$ (and hence $P$) contains both
literals of $C$. Clearly, $C$ is the required clause. $\blacksquare$

The soundness of Steps 5 and 6 of $\textsc{FindCS}$ is assumed in
the rest of the paper without explicit referring to Lemma
\ref{exclause}.

\begin{lemma} \label{validinput}
Let $(F,L,l,k)$ be a valid input and assume that
$Solve2ASLASAT(F,L,l,k)$ applies itself recursively. Then all the
children of $(F,L,l,k)$ in the search tree are valid inputs.
\end{lemma}

{\bf Proof.} Let $(F_1,L_1,l,k_1)$ be a child of $(F,L,l,k)$ .
Observe that $k_1 \geq k-1$. Observe also that $k>0$ because
$\textsc{FindCS}(F,L,l,k)$ would not apply itself recursively if
$k=0$. It follows that $k_1 \geq 0$.

It remains to prove that $(F_1,L_1,l)$ is a valid instance of the
\textsc{2-aslasat} problem. If $k_1=k-1$ then $(F_1,L_1,l)=(F
\setminus C,L,l)$ where $C$ is the clause selected on Steps 5 and 6.
In this case the validity of instance $(F \setminus C,L,l)$
immediately follows from the validity of $(F,L,l)$. Consider the
remaining case where $(F_1,L_1,l,k_1)=(F,L \cup \{l^*\},l,k)$ where
$l^*$ is a literal of the clause $C=(l_1 \vee l_2)$ selected on
Steps 5 and 6. In particular, we are going to show that

\begin{itemize}
\item $L \cup \{l^*\}$ is non-contradictory;
\item $Var(l) \notin Var(L \cup \{l^*\}$;
\item $SWRT(F, L \cup \{l^*\})$ is true.
\end{itemize}

That $L \cup \{l^*\}$ is non-contradictory follows from description
of the algorithm because it is explicitly stated that the literal
being joined to $L$ does not belong to $\neg(L \cup \{l\})$. This
also implies that the second condition may be violated only if
$l^*=l$. In this case assume that $C$ is selected on Step 5. Then
w.l.o.g. $l_1 \in \neg L$ and $l_2=l$. Let $P$ be a satisfying
assignment of $F$ which does not intersect with $\neg L$ (existing
since $(SWRT(F,L)$ is true). Then $l_2 \in P$, i.e. $SWRT(F,L \cup
\{l\})$ is true, which is impossible since in this the algorithm
would stop at Step 1. The assumption that $C$ is selected on Step 6
also leads to a contradiction because on the one hand $SWRT(F, l)$
is false by Lemma \ref{impl} due to existence of a walk from $\neg
l$ to $\neg l$, on the other hand $SWRT(F,l)$ is true by the
selection criterion. It follows that $Var(l) \notin Var(L \cup
\{l^*\})$.

Let us prove the last item. Assume first that $C$ is selected on
Step 5 and assume w.l.o.g. that $l_1 \in \neg L$. Then, by the first
statement, $l^*=l_2$. Moreover, as noted in the previous paragraph
$l_2 \in P$ where $P$ is a satisfying assignment of $F$ which does
intersect with $\neg L$, i.e. $SWRT(F, L \cup \{l_2\})$ is true in
the considered case. Assume that $C$ is selected on Step 6 and let
$w$ be the walk from $\neg l$ to $\neg l$ in $F$ to which $C$
belongs. Observe that $F$ has a walk $w'$ from $l^*$ to $\neg l$: if
$l^*$ is a first literal of $C$ w.r.t. $w$ then let $w'$ be a suffix
of $w$ whose first literal is $l^*$, otherwise let be the suffix of
$reverse(w)$ whose first literal is $l^*$. Assume that $SWRT(F, L
\cup \{l^*\})$ is false. Since $L \cup \{l^*\}$ is non-contradictory
by the first item, $Var(l^*) \notin Var(L)$. It follows that
$(F,L,l^*)$ is a valid instance of the \textsc{2-aslasat} problem.
In this case, by Theorem \ref{insatisf}, $F$ has either a walk from
$\neg L$ to $\neg l^*$ or a walk from $\neg l^*$ to $\neg l^*$. The
latter is ruled out by Lemma \ref{impl} because $SWRT(F,l^*)$ is
true by selection of $C$. Let $w''$ be a walk from $\neg L$ to $\neg
l^*$ in $F$. Then $w''+w'$ is a walk of $F$ from $\neg L$ to $\neg
l$ in contradiction to our assumption that $C$ is selected on Step
6. Thus $SWRT(F, L \cup \{l^*\})$ is true. The proof of the present
lemma is now complete. $\blacksquare$

Now we introduce two measures of the input of the $Solve2ASLASAT$
procedure. Let $\alpha(F,L,l,k)=|Var(F) \setminus Var(L)|+k$ and
$\beta(F,L,l,k)=max(0,2k-SepSize(F,\neg L,\neg l))$.

\begin{lemma} \label{alphameasure}
Let $(F,L,l,k)$ be a valid input and let $(F_1,L_1,l,k_1)$ be a
child of $(F,L,l,k)$. Then $\alpha(F,L,l,k)>\alpha(F_1,L_1,l,k_1)$.
\end{lemma}

{\bf Proof.} If $k_1=k-1$ then the statement is clear because the
first item in the definition of the $\alpha$-measure does not
increase and the second decreases. So, assume that
$(F_1,L_1,l,k_1)=(F,L \cup \{l^*\},l,k)$. In this case it is
sufficient to prove that $Var(l^*) \notin Var(L)$. Due to the
validity of $(F,L \cup \{l^*\},l,k)$ by Lemma \ref{validinput}, $l^*
\notin \neg L$, so it remains to prove that $l^* \notin L$. Assume
that  $l^* \in L$. Then the clause $C$ is selected on Step 6.
Indeed, if $C$ is selected on Step 5 then one of its literals
belongs to $\neg L$ and hence cannot belong to $L$, due to the
validity of $(F,L,l,k)$ (and hence being $L$ non-contradictory),
while the variable of the other literal does not belong to $Var(L)$
at all. Let $w$ be the walk from $\neg l$ to $\neg l$ in $F$ to
which $C$ belongs. Due to the validity of $(F,L \cup \{l^*\},l,k)$
by Lemma \ref{validinput}, $l^* \neq \neg l$. Therefore either $w$
or $reverse(w)$ has a suffix which is a walk from $\neg l^*$ to
$\neg l$, i.e. a walk from $\neg L$ to $\neg l$. But this
contradicts the selection of $C$ on Step 6. So, $l^* \notin L$ and
the proof of the lemma is complete. $\blacksquare$

For the next lemma we extend our terminology. We call a node
$(F',L',l,k')$ of $ST(F,L,l,k)$ a \emph{trivial} node if it is a
leaf or its only child is of the form $(F',L' \cup \{l^*\},l,k')$
for some literal $l^*$.

\begin{lemma} \label{betameasure}
Let $(F,L,l,k)$ be a valid input and let $(F_1,L_1,l,k_1)$ be a
child of $(F,L,l,k)$. Then $\beta(F,L,l,k) \geq
\beta(F_1,L_1,l,k_1)$. Moreover if $(F,L,l,k)$ is a non-trivial node
then $\beta(F,L,l,k)> \beta(F_1,L_1,l,k_1)$.
\end{lemma}

{\bf Proof.} Note that $\beta(F,L,l,k)>0$ because if
$\beta(F,L,l,k)=0$ then $\textsc{FindCS}(F,L,l,k)$ does not apply
itself recursively, i.e. does not have children. It follows that
$\beta(F,L,l,k)=2k-SepSize(F,\neg L, \neg l)>0$. Consequently, to
show that $\beta(F,L,l,k)>\beta(F_1,L_1,l,k_1)$ or that
$\beta(F,L,l,k) \geq \beta(F_1,L_1,l,k_1)$ it is sufficient to show
that $2k-SepSize(F,\neg L, \neg l)>2k_1-SepSize(F_1,\neg L_1, \neg
l)$ or $2k-SepSize(F,\neg L, \neg l) \geq 2k_1-SepSize(F_1,\neg L_1,
\neg l)$, respectively.

Assume first that $(F_1,L_1,l,k_1)=(F \setminus C,L,l,k-1)$. Observe
that $SepSize(F \setminus C,\neg L,\neg l) \geq SepSize(F,\neg
L,\neg l)-1$. Indeed assume the opposite and let $S$ be a separator
w.r.t. to $\neg L$ and $\neg l$ in $F \setminus C$ whose size is at
most $SepSize(F,\neg L,\neg l)-2$. Then $S \cup \{C\}$ is a
separator w.r.t. $\neg L$ and $\neg l$ in $F$ of size at most
$SepSize(F,\neg L,\neg l)-1$ in contradiction to the definition of
$SepSize(F,\neg L,\neg l)$. Thus $2(k-1)-SepSize(F \setminus C,\neg
L,\neg l)=2k-SepSize(F \setminus C,\neg L,\neg l)-2 \leq
2k-SepSize(F,\neg L,\neg l)-1<2k-SepSize(F,\neg L,\neg l)$.

Assume now that $(F_1,L_1,l,k_1)=(F,L \cup \{l^*\},l,k)$ for some
literal $l^*$. Clearly, $SepSize(F, \neg L,\neg l) \leq SepSize(F,
\neg (L \cup \{l^*\}), \neg l)$ due to being $\neg L$ a subset of
$\neg (L \cup \{l^*\})$. It follows that $2k-SepSize(F, \neg L,\neg
l) \geq 2k-SepSize(F, \neg (L \cup \{l^*\}), \neg l)$. It remains to
show that $\geq$ can be replaced by $>$ in case where $(F,L,l,k)$ is
a non-trivial node. It is sufficient to show that in this case
$SepSize(F,\neg L,\neg l)<SepSize(F,\neg (L \cup \{l^*\}),\neg l)$.
If $(F,L,l,k)$ is a non-trivial node then the recursive call
$\textsc{FindCS}(F,L \cup \{l^*\},l,k)$ is applied on Steps 8.2,
8.4, or 9.3. In the last case, it is explicitly said that $l^*$ is
not a neutral literal in $(F,L,l)$. Consequently, $SepSize(F,\neg
L,\neg l)<SepSize(F,\neg (L \cup \{l^*\}),\neg l)$ by definition.

For the first two cases note that Step 8 is applied only if the
clause $C$ is selected on Step 6. That is, $F$ has no walk from
$\neg L$ to $\neg l$. In particular, $F$ has no path from $\neg L$
to $\neg l$, i.e. $SepSize(\neg L,\neg l)=0$. Let $w$ be the walk
from $\neg l$ to $\neg l$ in $F$ to which $C$ belongs. Note that by
Lemma \ref{validinput}, $(F,L \cup \{l^*\},l,k)$ is a valid input,
in particular $Var(l^*) \neq Var(l)$. Therefore either $w$ or
$reverse(w)$ has a suffix which is a walk from $\neg l^*$ to $\neg
l$, i.e. a walk from $\neg (L \cup \{l^*\})$ to $\neg l$. Applying
Lemma \ref{walkpath} together with Lemma \ref{validinput}, we see
that $F$ has a path from $\neg (L \cup \{l^*\})$ to $\neg l$, i.e.
$SepSize(F, \neg (L \cup \{l^*\},\neg l)>0$. $\blacksquare$

\begin{lemma} \label{summary}
Let $(F,L,l,k)$ be a valid input. Then the following statements are
true regarding $ST(F,L,l,k)$.
\begin{itemize}
\item The height of $ST(F,L,l,k)$ is at most $\alpha(F,L,l,k)$.
\footnote{Besides providing the upper bound on the height of
$ST(F,L,l,k)$, this statement claims that $ST(F,L,l,k)$ is finite
and hence we may safely refer to a path between two nodes.}
\item Each node $(F',L',l,k')$ of $ST(F,L,l,k)$ is a valid input,
the subtree rooted by $(F',L',l,k')$ is $ST(F',L',l,k')$ and
$\alpha(F',L',l,k')<\alpha(F,L,l,k)$.
\item For each node $(F',L',l,k')$ of $ST(F,L,l,k)$,
$\beta(F',L',l,k') \leq \beta(F,L,l,k)-t$ where $t$ is the number of
non-trivial nodes besides $(F',L',l,k')$ in the path from
$(F,L,l,k)$ to $(F',L',l,k')$ of $ST(F,L,l,k)$.
\end{itemize}
\end{lemma}

{\bf Proof.} This lemma is clearly true if $(F,L,l,k)$ has no
children. Consequently, it is true if $\alpha(F,L,l,k)=0$. Now,
apply induction on the size of $\alpha(F,L,l,k)$ and assume that
$\alpha(F,L,l,k)>0$. By the induction assumption, Lemma
\ref{validinput}, and Lemma \ref{alphameasure}, the present lemma is
true for any child of $(F,L,l,k)$. Consequently, for any child
$(F^*,L^*,l,k^*)$ of $(F,L,l,k)$, the height of $ST(F^*,L^*,l,k^*)$
is at most $\alpha(F^*,L^*,l,k^*)$. Hence the first statement
follows by Lemma \ref{alphameasure}. Furthermore, any node
$(F',L',l,k')$ of $ST(F,L,l,k)$ belongs to $ST(F^*,L^*,l,k^*)$ of
some child $(F^*,L^*,l,k^*)$ of $(F,L,l,k)$ and the subtree rooted
by $(F',L',l,k')$ in $ST(F,L,l,k)$ is the subtree rooted by
$(F',L',l,k')$ in $ST(F^*,L^*,l,k^*)$. Consequently, $(F',L',l,k')$
is a valid input, the subtree rooted by it is $ST(F',L',l,k')$, and
$\alpha(F',L',l,k') \leq \alpha(F^*,L^*,l,k^*)<\alpha(F,L,l,k)$, the
last inequality follows from Lemma \ref{alphameasure}. Finally,
$\beta(F',L',l,k') \leq \beta(F^*,L^*,l,k^*)-t^*$ where $t^*$ is the
number of non-trivial nodes besides $(F',L',l,k')$ in the path from
$(F^*,L^*,l,k^*)$ to $(F',L',l,k')$ in $ST(F^*,L^*,l,k^*)$, and
hence in $ST(F,L,l,k)$ \footnote{Note that this inequality applies
to the case where $(F',L',l,k')=(F^*,L^*,l,k^*)$.}. If $(F,L,l,k)$
is a trivial node then $t=t^*$ and the last statement of the present
lemma is true by Lemma \ref{betameasure}. Otherwise $t=t^*+1$ and by
another application of Lemma \ref{betameasure} we get that
$\beta(F',L',l,k') \leq \beta(F,L,l,k)-t^*-1=\beta(F,L,l,k)-t$.
$\blacksquare$

\subsection{Correctness Proof}
\begin{theorem} \label{correct}
Let $(F,L,l,k)$ be a valid input. Then $\textsc{FindCS}(F,L,l,k)$
correctly solves the parameterized \textsc{2-aslasat} problem. That
is, if $\textsc{FindCS}(F,L,l,k)$ returns a set, this set is a CS of
$(F,L,l)$ of size at most $k$. If $\textsc{FindCS}(F,L,l,k)$ returns
'NO' then $(F,L,l)$ has no CS of size at most $k$.
\end{theorem}

{\bf Proof.} Let us prove first the correctness of
$\textsc{FindCS}(F,L,l,k)$ for the cases when the procedure does not
apply itself recursively. It is only possible when the procedure
returns an answer on Steps 1-4. If the answer is returned on Step 1
then the validity is clear because nothing has to be removed from
$F$ to make it satisfiable w.r.t. $L$ and $l$. If the answer is
returned on Step 2 then $SWRT(F,L \cup \{l\})$ is false (since the
condition of Step 1 is not satisfied) and consequently the size of a
CS of $(F,L,l)$ is at least 1. On the other hand, $k=0$ and hence
the answer 'NO' is valid in the considered case. For the answer
returned on Step 3 observe that $Clauses(F)$ is clearly a CS of
$(F,L,l)$ (since $SWRT(\emptyset,L \cup \{l\})$ is true) and the
size of $Clauses(F)$ does not exceed $k$ by the condition of Step 3.
Therefore the answer returned on this step is valid. Finally if the
answer is returned on Step 4 then the condition of Step 4 is
satisfied. According to Corollary \ref{lowerbound}, this condition
implies that any CS of $(F,L,l)$ has the size greater than $k$,
which justifies the answer 'NO' in the considered step.

Now we prove correctness of $\textsc{FindCS}(F,L,l,k)$ by induction
on $\alpha(F,L,l,k)$. Assume first that $\alpha(F,L,l,k)=0$. Then it
follows that $k=0$ and, consequently, $\textsc{FindCS}(F,L,l,k)$
does not apply itself recursively (the output is returned on Step 1
or Step 2). Therefore, the correctness of $\textsc{FindCS}(F,L,l,k)$
follows from the previous paragraph. Assume now that
$\alpha(F,L,l,k)>0$ and that the theorem holds for any valid input
$(F',L',l,k')$ such that $\alpha(F',L',l,k')<\alpha(F,L,l,k)$. Due
to the previous paragraph we may assume that
$\textsc{FindCS}(F,L,l,k)$ applies itself recursively, i.e. the node
$(F,L,l,k)$ has children in $ST(F,L,l,k)$.

\begin{cclaim} \label{children}
Let $(F_1,l_1,l,k_1)$ be a child of $(F,L,l,k)$. Then
$\textsc{FindCS}(F_1,L_1,l,k_1)$ is correct.
\end{cclaim}

{\bf Proof.} By Lemma \ref{validinput}, $(F_1,L_1,l,k_1)$ is a valid
input. By Lemma \ref{alphameasure},
$\alpha(F_1,L_1,l,k_1)<\alpha(F,L,l,k)$. The claim follows by the
induction assumption. $\square$

Assume that $\textsc{FindCS}(F,L,l,k)$ returns a set $S$. By
description of the algorithm, either $S$ is returned by
$\textsc{FindCS}(F,L \cup \{l^*\},l,k)$ for a child $(F,L \cup
\{l^*\},l,k)$ of $(F,L,l,k)$ or $S=S_1 \cup \{C\}$ and $S_1$ is
returned by $\textsc{FindCS}(F \setminus C,L,l,k-1)$ for a child $(F
\setminus C,L,l,k-1)$ of $(F,L,l,k)$. In the former case, the
validity of output follows from Claim \ref{children} and from the
easy observation that a CS of $(F,L \cup \{l^*\},l,k)$ is a CS of
$(F,L,l,k)$ because $L$ is a subset of $L \cup \{l^*\}$. In the
latter case, it follows from Claim \ref{children} that $|S_1| \leq
k-1$ and that $S_1$ is a CS of $(F \setminus C,L,l)$ i.e. $SWRT((F
\setminus C) \setminus S_1, L \cup \{l\})$ is true. But $(F
\setminus C) \setminus S_1=F \setminus (S_1 \cup \{C\})=F \setminus
S$. Consequently $S$ is a CS of $(F,L,l)$ of size at most $k$, hence
the output is valid in the considered case.

Consider now the case where $\textsc{FindCS}(F,L,l,k)$ returns 'NO'
and assume by contradiction that there is a CS $S$ of $(F,L,l)$ of
size at most $k$. Assume first that 'NO' is returned on Step 7.3. It
follows that $C \notin S$ because otherwise $S \setminus C$ is a CS
of $(F \setminus C,L,l)$ of size at most $k-1$ and hence, by Claim
\ref{children}, the recursive call of Step 7.2. would not return
'NO'. However, this means that any satisfying assignment of $F
\setminus S$ which does not intersect with $\neg (L \cup \{l\})$
(which exists by definition) cannot satisfy clause $C$, a
contradiction. Assume now that 'NO' is returned on Step 10. By Claim
\ref{children}, $(F,L \cup \{l_2\},l)$ has no CS of size at most
$k$. Therefore, according to Theorem \ref{pseudo}, the size of a SCS
of $(F,L,l)$ is at least $k+1$ which contradicts the existence of
$S$. Finally assume that 'NO' is returned on Step 8.7. or on Step
9.5. Assume first that the clause $C$ selected on Steps 5 and 6 does
not belong to $S$. Let $P$ be a satisfying assignment of $(F
\setminus S)$ which does not intersect with $\neg(L \cup \{l\})$.
Then at least one literal $l^*$ of $C$ is contained in $P$. This
literal does not belong to $\neg( L \cup \{l\})$ and hence
$\textsc{FindCS}(F,L \cup \{l^*\},l,k)$ has been applied and
returned 'NO'. However, $P$ witnesses that $S$ is a CS of $(F, L
\cup \{l^*\},l,k)$ of size at most $k$, that is $\textsc{FindCS}(F,L
\cup \{l^*\},l,k)$ returned an incorrect answer in contradiction to
Claim \ref{children}. Finally assume that $C \in S$. Then $S
\setminus C$ is a CS of $(F \setminus C,L,l)$ of size at most $k-1$
and hence answer 'NO' returned by $\textsc{FindCS}(F \setminus
C,L,l)$ contradicts Claim \ref{children}. Thus the answer 'NO'
returned by $\textsc{FindCS}(F,L,l,k)$ is valid. $\blacksquare$

\subsection{Evaluation of the runtime.}

\begin{theorem} \label{numleaves}
Let $(F,L,l,k)$ be a valid input. Then the number of leaves of
$ST(F,L,l,k)$ is at most $\sqrt{5}^t$, where $t=\beta(F,L,l,k)$.
\end{theorem}

{\bf Proof of Theorem \ref{numleaves}} Since $\beta(F,L,l,k) \geq 0$
by definition, $\sqrt{5}^t \geq 1$. Hence if
$\textsc{FindCS}(F,L,l,k)$ does not apply itself recursively, i.e.
$ST(F,L,l,k)$ has only one node, the theorem clearly holds. We prove
the theorem by induction on $\alpha(F,L,l,k)$. If
$\alpha(F,L,l,k)=0$ then as we have shown in the proof of Theorem
\ref{correct}, $\textsc{FindCS}(F,L,l,k)$ does not apply itself
recursively and hence the theorem holds as shown above. Assume that
$\alpha(F,L,l,k)>0$ and that the theorem holds for any valid input
$(F',L',l,k')$ such that $\alpha(F',L',l,k')<\alpha(F,L,l,k)$.
Clearly we may assume that $(F,L,l,k)$ applies itself recursively
i.e. $ST(F,L,l,k)$ has more than 1 node.

\begin{cclaim} \label{children2}
For any non-root node $(F',L',l,k')$ of $ST(F,L,l,k)$, the subtree
of $ST(F,L,l,k)$ rooted by $(F',L',l,k')$ has at most
$\sqrt{5}^{t'}$ leaves, where $t'=\beta(F',L',l,k')$.
\end{cclaim}

{\bf Proof.} According to Lemma \ref{summary}, $(F',L',l,k')$ is a
valid input, $\alpha(F',L',l,k')<\alpha(F,L,l,k)$, and the subtree
of $ST(F,L,l,k)$ rooted by $(F',L',l,k')$ is $ST(F',L',l,k')$.
Therefore the claim follows by the induction assumption. $\square$

If $(F,L,l,k)$ has only one child $(F_1,L_1,l,k_1)$ then clearly the
number of leaves of $ST(F,L,l,k)$ equals the number of leaves of the
subtree rooted by $(F_1,L_1,l,k_1)$ which, by Claim \ref{children2},
is at most $\sqrt{5}^{t_1}$, where $t_1=\beta(F_1,L_1,l,k_1)$.
According to Lemma \ref{betameasure}, $t_1 \leq t$ so the present
theorem holds for the considered case. If $(F,L,l,k)$ has 2 children
$(F_1,L_1,l,k_1)$ and $(F_2,L_2,l,k_2)$ then the number of leaves of
$ST(F,L,l,k)$ is the sum of the numbers of leaves of subtrees rooted
by $(F_1,L_1,l,k_1)$ and $(F_2,L_2,l,k_2)$ which, by Claim
\ref{children2}, is at most $\sqrt{5}^{t_1}+\sqrt{5}^{t_2}$, where
$t_i=\beta(F_i,L_i,l,k_i)$ for $i=1,2$. Taking into account that
$(F,L,l,k)$ is a non-trivial node and applying Lemma
\ref{betameasure}, we get that $t_1<t$ and $t_2<t$. hence the number
of leaves of $ST(F,L,l,k)$ is at most
$(2/\sqrt{5})*(\sqrt{5}^{t})<\sqrt{5}^{t}$, so the theorem holds for
the considered case as well.

For the case where $(F,L,l,k)$ has 3 children, denote them by
$(F_i,L_i,l,k_i)$, $i=1,2,3$. Assume w.l.o.g. that
$(F_1,L_1,l,k_1)=(F, L \cup \{l_1\},l,k)$, $(F_2,L_2,l,k_2)=(F,L
\cup \{l_2\},l,k)$, $(F_3,L_3,l,k_3)=(F\setminus C,l,k-1)$, where
$C=(l_1 \vee l_2)$ is the clause selected on steps 5 and 6. Let
$t_i=\beta(F_i,L_i,l,k_i)$ for $i=1,2,3$.

\begin{cclaim} \label{tlast}
$t \geq 2$ and $t_3 \leq t-2$.
\end{cclaim}

{\bf Proof.} Note that $k>0$ because otherwise
$\textsc{FindCS}(F,L,l,k)$ does not apply itself recursively.
Observe also that $SepSize(F,\neg L,\neg l)=0$ because clause $C$
can be selected only on Step 6, which means that $F$ has no walk
from $\neg L$ to $\neg l$ and, in particular, $F$ has no path from
$\neg L$ to $\neg l$. Therefore $2k-Sepsize(F,\neg L,\neg l)=2k \geq
2$ and hence $t=\beta(F,L,l,k)=2k \geq 2$. If $t_3=0$ the second
statement of the claim is clear. Otherwise $t_3=2(k-1)-SepSize(F
\setminus (l_1 \vee l_2),\neg L,\neg l)=2(k-1)-0=2k-2=t-2$.
$\square$

Assume that some $ST(F_i,L_i,l,k_i)$ for $i=1,2$ has only one leaf.
Assume w.l.o.g. that this is $ST(F_1,L_1,l,k_1)$. Then the number of
leaves of $ST(F,L,l,k)$ is the sum of the numbers of leaves of the
subtrees rooted by $(F_2,L_2,l,k_2)$ and $(F_3,L_3,l,k_3)$ plus one.
By Claims \ref{children2} and \ref{tlast}, and Lemma
\ref{betameasure}, this is at most
$\sqrt{5}^{t-1}+\sqrt{5}^{t-2}+1$. Then
$\sqrt{5}^t-\sqrt{5}^{t-1}-\sqrt{5}^{t-2}-1 \geq
\sqrt{5}^{2}-\sqrt{5}^{2-1}-\sqrt{5}^{2-2}-1=5-\sqrt{5}-2>0$, the
first inequality follows from Claim \ref{tlast}. That is, the
present theorem holds for the considered case.

It remains to assume that both $ST(F_1,L_1,l,k_1)$ and
$ST(F_2,L_2,l,k_2)$ have at least two leaves. Then for $i=1,2$,
$ST(F_i,L_i,l,k_i)$ has a node having at least two children. Let
$(FF_i,LL_i,l,kk_i)$ be such a node of $ST(F_i,L_i,l,k_i)$ which
lies at \emph{the smallest distance} from $(F,L,l,k)$ in
$ST(F,L,l,k)$.

\begin{cclaim} \label{descleaves}
The number of leaves of the subtree rooted by $(FF_i,LL_i,l,kk_i)$
is at most $(2/5)*\sqrt{5}^{t}$.
\end{cclaim}

{\bf Proof.} Assume that $(FF_i,LL_i,l,kk_i)$ has 2 children and
denote them by $(FF^*_1,LL^*_1,l,kk^*_1)$ and
$(FF^*_2,LL^*_2,l,kk^*_2)$. Then the number of leaves of the subtree
rooted by  $(FF_i,LL_i,l,kk_i)$ equals the sum of numbers of leaves
of the subtrees rooted by $(FF^*_1,LL^*_1,l,kk^*_1)$ and
$(FF^*_2,LL^*_2,l,kk^*_2)$. By Claim \ref{children2}, this sum does
not exceed $2*\sqrt{5}^{t^*}$ where $t^*$ is the maximum of
$\beta(FF^*_j,LL^*_j,l,kk^*_j)$ for $j=1,2$. Note that the path from
$(F,L,l,k)$ to any $(FF^*_j,LL^*_j,l,kk^*_j)$ includes at least 2
non-trivial nodes besides $(FF^*_j,LL^*_j,l,kk^*_j)$, namely
$(F,L,l,k)$ and $(FF_i,LL_i,l,kk_i)$. Consequently, $t^* \leq t-2$
by Lemma \ref{summary} and the present claim follows for the
considered case.

Assume that $(FF_i,LL_i,l,kk_i)$ has 3 children. Then let
$tt_i=\beta(FF_i,LL_i,l,kk_i)$ and note that according to Claim
\ref{children2}, the number of leaves of the subtree rooted by
$(FF_i,LL_i,l,kk_i)$ is at most $\sqrt{5}^{tt_i}$. Taking into
account that $(FF_i,LL_i,l,kk_i)$ is a valid input by Lemma
\ref{summary} and arguing analogously to the second sentence of the
proof of Claim \ref{tlast}, we see that $SepSize(FF_i, \neg LL_i,
\neg l)=0$. On the other hand, using the argumentation in the last
paragraph of the proof of Lemma \ref{betameasure}, we can see that
$SepSize(F_i, \neg L_i,l)>0$. This means that $(F_i,L_i,l,k_i) \neq
(FF_i,LL_i,l,kk_i)$. Moreover, the path from $(F_i,L_i,l,k_i)$ to
$(FF_i,LL_i,l,kk_i)$ includes a pair of consecutive nodes
$(F',L',l,k')$ and $(F'',L'',l,k'')$, being the former the parent of
the latter, such that $SepSize(F',\neg L',\neg l)>SepSize(F'', \neg
L'',\neg l)$. This only can happen if $k''=k'-1$ (for otherwise
$(F'',L'',l,k'')=(F',L' \cup \{l'\},l,k')$ for some literal $l'$ and
clearly adding a literal to $L'$ does not decrease the size of the
separator). Consequently, $(F',L',l,k')$ is a non-trivial node.
Therefore, the path from $(F,L,l,k)$ to $(FF_i,LL_i,l,kk_i)$
includes at least 2 non-trivial nodes besides $(FF_i,LL_i,l,kk_i)$:
$(F,L,l,k)$ and $(F',L',l,k')$. That is $tt_i \leq t-2$ by Lemma
\ref{summary} and the present claims follows for this case as well
which completes its proof. $\square$

It remains to notice that the number of leaves of $ST(F,L,l,k)$ is
the sum of the numbers of leaves of subtrees rooted by
$(FF_1,LL_1,l,kk_1)$, $(FF_2,LL_2,l,kk_2)$, and $(F_3,L_3,l,k_3)$
which, according to Claims \ref{children2} \ref{tlast} and
\ref{descleaves}, is at most $5*\sqrt{5}^{t-2}=\sqrt{5}^t$.
$\blacksquare$

\begin{theorem} \label{2aslasat}
Let $(F,L,l,k)$ be an instance of the parameterized
\textsc{2-aslasat} problem. Then the problem can be solved in time
$O(5^k*k(n+k)*(m+|L|))$, where $n=|Var(F)|$, $m=|Clauses(F)|$.
\end{theorem}

{\bf Proof.} According to assumptions of the theorem, $(F,L,l,k)$ is
a valid input. Assume that $F$ is represented by its implication
graph $D=(V,A)$ which is almost identical to the implication graph
of $F$ with the only difference that $V(D)$ corresponds to $Var(F)
\cup Var(L) \cup Var(l')$, that is if for any literal $l'$ such that
$Var(l') \in (Var(L) \cup \{Var(l)\}) \setminus Var(F)$, $D$ has
isolated nodes corresponding to $l'$ and $\neg l'$. We also assume
that the nodes corresponding to $L$, $\neg L$, $l$, $\neg l$ are
specifically marked. This representation of $(F,L,l,k)$ can be
obtained in a polynomial time from any other reasonable
representation. It follows from Theorem \ref{correct} that
$\textsc{FindCS}(F,L,l,k)$ correctly solves the parameterized
\textsc{2-aslasat} problem with respect to the given input. Let us
evaluate the complexity of $\textsc{FindCS}(F,L,l,k)$. According to
Lemma \ref{summary}, the height of the search tree is at most
$\alpha(F,L,l,k) \leq n+k$. Theorem \ref{numleaves} states that the
number of leaves of $ST(F,L,l,k)$ is at most $\sqrt{5}^{t}$ where
$t=\beta(F,L,l,k)$. Taking into account that $t \leq 2k$, the number
of leaves of $ST(F,L,l,k)$ is at most $5^k$. Consequently, the
number of nodes of the search tree is at most $5^k*(n+k)$. The
complexity of $\textsc{FindCS}(F,L,l,k)$ can be represented as the
number of nodes multiplied by the complexity of the operations
performed \emph{within} the given recursive call.

Let us evaluate the complexity of $\textsc{FindCS}(F,L,l,k)$ without
taking into account the complexity of the subsequent recursive
calls. First of all note that each literal of $F$ belongs to a
clause and each clause contains at most 2 distinct literals.
Consequently, the number of clauses of $F$ is at least half of the
number of literals of $F$ and, as a result, at least half of the
number of variables. This notice is important because most of
operations of $\textsc{FindCS}(F,L,l,k)$ involve doing Depth-First
Search (DFS) or Breadth-First Search (BFS) on graph $D$, which take
$O(V+A)$. In our case $|V|=O(n+|L|)$ and $|A|=O(m)$. Since $n=O(m)$,
$O(V+A)$ can be replaced by $O(m+|L|)$.

The first operation performed by $\textsc{FindCS}(F,L,l,k)$ is
checking whether $SWRT(F,L \cup \{l\})$ is true. Note that this is
equivalent to checking the satisfiability of a \textsc{2-cnf} $F'$
which is obtained from $F$ by adding clauses $(l' \vee l')$ for each
$l' \in L \cup \{l\}$. It is well known \cite{Papa} that the given
\textsc{2-cnf} formula $F'$ is \emph{not} satisfiable if and only if
there are literals $l'$ and $\neg l'$ which belong to the same
strongly connected component of the implication graph of $F'$. The
implication graph $D'$ of $F'$ can be obtained from $D$ by adding
arcs that correspond to the additional clauses. The resulting graph
has $O(m+|L|)$ vertices and $O(m+|L|)$ arcs. The partition into the
strongly connected components can be done by a constant number of
applications of the DFS algorithm. Hence the whole Step 1 takes
$O(m+|L|)$. Steps 2 and 3 take $O(1)$. According to Theorem
\ref{genmenger}, Step 4 can be performed by assigning all the arcs
of $D$ a unit flow, contracting all the vertices of $L$ into a
source $s$, identifying $\neg l$ with the sink $t$, and checking
whether $k+1$ units of flow can be delivered from $s$ to $t$. This
can be done by $O(k)$ iterations of the Ford-Fulkerson algorithm,
where each iteration is a run of BFS and hence can be performed on
$O(m+|L|)$. Consequently, Step 4 can be performed in $O((m+|L|)*k)$.
Checking the condition of Step 5 can be done by BFS and hence takes
$O(m+|L|)$. Moreover, if the required walk exists, BFS finds the
shortest one and, as noted in the proof of Lemma \ref{exclause}, a
required clause is the first clause of this walk. Hence, the whole
Step 5 can be performed in $O(m+|L|)$. The proof of Lemma
\ref{exclause} also outlines an algorithm implementing Step 6:
choose an arbitrary walk $w$ from $\neg l$ to $\neg l$ in $F$,
(which, as noted in the proof of Theorem \ref{genmenger},
corresponds to a walk from $l$ to $\neg l$ in $D$), find a
satisfying assignment $P$ of $F$ which does not intersect with $\neg
L$ and choose a clause of $w$ whose both literals are satisfied by
$P$. Taking into account the above discussion, all the operations
take $O(m+|L|)$, hence Step 6 takes this time. Note that preparing
an input for a recursive call takes $O(1)$ because this preparation
includes removal of one clause from $F$ or adding one literal to $L$
(with introducing appropriate changes to the implication graph).
Therefore Steps 7 and 8 take $O(1)$. Step 9 takes $O((m+|L|)*k)$ on
the account of neutrality checking: $O(k)$ iterations of the
Ford-Fulkerson algorithm are sufficient because $SepSize(F, \neg L,
\neg l) \leq k$ due to insatisfaction of the condition of Step 4.
Step 10 takes $O(1)$ on the account of input preparation for the
recursive call. Thus the complexity of processing $(F,L,l,k)$ is
$O((m+|L|)*k)$.

Finally, note that for any subsequent recursive call $(F',L',l,k')$
the implication graph of $(F',L',l)$ is a subgraph of the graph of
$(F,L,l)$: every change of graph in the path from $(F,L,l,k)$ to
$(F',L',l,k')$ is caused by removal of a clause or adding to the
second parameter a literal of a variable of $F$. Consequently, the
complexity of any recursive call is $O((m+|L|)*k)$ and the time
taken by the entire run of $\textsc{FindCS}(F,L,l,k)$ is
$O(5^k*k(n+k)*(m+|L|))$ as required. $\blacksquare$

\section{Fixed-Parameter Tractability of 2-ASAT problem}
In this section we prove the main result of the paper,
fixed-parameter tractability of the 2-ASAT problem.

\begin{theorem} \label{2asatfpt}
The \textsc{2-asat} problem with input $(F,k)$ where $F$ is a
\textsc{2-cnf} formula with possible repeated occurrences of
clauses, can be solved in $O(15^k*k*m^3)$, where $m$ is the number
of clauses of $F$.
\end{theorem}

{\bf Proof.} We introduce the following 2 intermediate problems.
\begin{quote}
  \noindent{\bfseries Problem I1}\\
  \emph{Input:} A satisfiable \textsc{2-cnf} formula $F$, a non-contradictory
  set of literals $L$, a parameter $k$\\
  \emph{Output:} A set $S \subseteq Clauses(F)$ such that $|S| \leq k$
  and $SWRT(F \setminus S, L)$ is true, if there is such a set $S$; 'NO'
  otherwise.
\end{quote}

\begin{quote}
  \noindent{\bfseries Problem I2}\\
  \emph{Input:} A \textsc{2-cnf} formula $F$, a parameter $k$, and a set
  $S \subseteq Clauses(F)$ such $|S|=k+1$ and
  $F \setminus S$ is satisfiable\\
  \emph{Output:} A set $Y \subseteq Clauses(F)$ such that $|Y| <
  |S|$ and $F \setminus Y$ is satisfiable, if there is such a set
  $Y$; 'NO' otherwise.
\end{quote}

The following two claims prove the fixed-parameter tractability of
Problem I1 through transformation of its instance into an instance
of \textsc{2-aslasat} problem and of Problem I2 through
transformation of its instance into an instance of Problem I1. Then
we will show that the \textsc{2-asat} problem with no repeated
occurrence of clauses can be solved through transformation of its
instance into an instance of Problem I2. Finally, we show that the
\textsc{2-asat} problem with repeated occurrences of clauses is
\textsc{fpt} through transformation of its instance into an instance
of \textsc{2-asat} without repeated occurrences of clauses.

%Add something about the implication graph
\begin{cclaim} \label{problemi1}
Problem I1 with the input $(F,L,k)$ can be solved in $O(5^k*k*m^2)$,
where, $m=|Clauses(F)|$.
\end{cclaim}

{\bf Proof.} Observe that we may assume that $Var(L) \subseteq
Var(F)$. Otherwise we can take a subset $L'$ such that
$Var(L')=Var(F) \cap Var(L)$ and solve problem I1 w.r.t. the
instance $(F,L',k)$. It is not hard to see that the resulting
solution applies to $(F,L,k)$ as well.

Let $P$ be a satisfying assignment of $F$. If $L \subseteq P$ then
the empty set can be immediately returned. Otherwise partition $L$
into two subsets $L_1$ and $L_2$ such that $L_1 \subseteq P$ and
$\neg L_2 \subseteq P$.

We apply a two stages transformation of formula $F$. On the first
stage we assign each clause of $F$ a unique index from $1$ to $m$,
introduce new literals $l_1, \dots, l_m$ of distinct variables which
do not intersect with $Var(F)$, and replace the $i$-th clause $(l'
\vee l'')$ by two clauses $(l' \vee l_i)$ and $(\neg l_i \vee l'')$.
Denote the resulting formula by $F'$. On the second stage we
introduce two new literals $l^*_1$ and $l^*_2$ such that $Var(l^*_1)
\notin Var(F')$,  $Var(l^*_2) \notin Var(F')$, and $Var(l^*_1) \neq
Var(l^*_2)$. Then we replace in the clauses of $F'$ each occurrence
of a literal of $L_1$ by $l^*_1$, each occurrence of a literal of
$\neg L_1$ by $\neg l^*_1$, each occurrence of a literal of $L_2$ by
$l^*_2$, and each occurrence of a literal of $\neg L_2$ by $\neg
l^*_2$. Let $F^*$ be the resulting formula.

We claim that $(F^*,\{l^*_1\},l^*_2)$ is a valid instance of the
\textsc{2-aslasat} problem. To show this we have to demonstrate that
all the clauses of $F^*$ are pairwise different and that
$SWRT(F^*,l^*_1)$ is true.

For the former, notice that all the clauses of $F^*$ are pairwise
different because each clause is associated with the unique literal
$l_i$ or $\neg l_i$. This also allows us to introduce new notation.
In particular, we denote the clause of $F^*$ containing $l_i$ by
$C(l_i)$ and the clause containing $\neg l_i$ by $C(\neg l_i)$.

For the latter let $P^*$ be a set of literals obtained from $P$ by
replacing $L_1$ by $l^*_1$ and $\neg L_2$ by $\neg l^*_2$. Observe
that for each $i$, $P^*$ satisfies either $C(l_i)$ or $C (\neg
l_i)$. Indeed, let $(l' \vee l'')$ be the \emph{origin} of $C(l_i)$
and $C(\neg l_i)$ i.e. the clause which is transformed into $(l'
\vee l_i)$ and $(\neg l_i \vee l'')$ in $F'$, then $(l' \vee l_i)$
and $(\neg l_i \vee l'')$ become respectively $C(l_i)$ and $C(\neg
l_i)$ in $F^*$ (with possible replacement of $l'$ or $l''$ or both).
Since $P$ is a satisfying assignment of $F$, $l' \in P$ or $l'' \in
P$. Assume the former. Then if $C(l_i)=(l' \vee l_i)$, $l' \in P^*$.
Otherwise, $l' \in L_1$ or $l' \in \neg L_2$. In the former case
$C(l_i)=(l^*_1 \vee l_i)$ and $l^*_1 \in P^*$ by definition; in the
latter case $C(l_i)=(\neg l^*_2 \vee l_i)$ and $\neg l^*_2 \in P^*$
by definition. So, in all the cases $P^*$ satisfies $C(l_i)$. It can
be shown analogously that if $l'' \in P$ then $P^*$ satisfies
$C(\neg l_i)$. Now, let $P^*_2$ be a set of literals which includes
$P^*$ and for each $i$ exactly one of $\{l_i \neg l_i\}$ selected as
follows. If $P^*$ satisfies $C(l_i)$ then $\neg l_i \in P^*_2$.
Otherwise $l_i \in P^*_2$. Thus $P^*_2$ satisfies all the clauses of
$F^*$. By definition $l^*_1 \in P^* \subseteq P^*_2$. It is also not
hard to show that $P^*_2$ is non-contradictory and that
$Var(P^*_2)=Var(F^*)$. Thus $P^*_2$ is a satisfying assignment of
$F^*$ containing $l^*_1$ which witnesses $SWRT(F^*,l^*_1)$ is true.

We are going to show that there is a set $S \subseteq Clauses(F)$
such that $|S| \leq k$ and $SWRT(F \setminus S,L)$ is true if and
only if $(F^*,\{l^*_1\},l^*_2)$ has a CS of size at most $k$.

Assume that there is a set $S$ as above. Let $S^* \subseteq
Clauses(F^*)$ be the set consisting of all clauses $C(l_i)$ such
that the clause with index $i$ belongs to $S$. It is clear that
$|S^*|=|S|$. Let us show that $S^*$ is a CS of
$(F^*,\{l^*_1\},l^*_2)$. Let $P$ be a satisfying assignment of $F
\setminus S$ which does not intersect with $\neg L$. Let $P_1$ be
the set of literals obtained from $P$ by replacing the set of all
the occurrences of literals of $L_1$ by $l^*_1$ and the set of all
the occurrences of literals of $L_2$ by $l^*_2$.

Observe that for each $i$, at least one of $\{C(l_i), C(\neg l_i)\}$
either belongs to $S^*$ or is satisfied by $P_1$. In particular,
assume that for some $i$, $C(l_i) \notin S^*$. Then the origin of
$C(l_i)$ and $C(\neg l_i)$ belongs to $F \setminus S$ and it can be
shown that $P_1$ satisfies $C(l_i)$ or $C(\neg l_i)$ similarly to
the way we have shown that $P^*$ satisfies $C(l_i)$ or $C(\neg l_i)$
three paragraphs above.

For each $i$, add to $P_1$ an appropriate $l_i$ or $\neg l_i$ so
that the remaining clauses of $F^* \setminus S^*$ are satisfied, let
$P_2$ be the resulting set of literals. Add to $P_2$ one arbitrary
literal of each variable of $Var(F^* \setminus S^*) \setminus
Var(P_2)$. It is not hard to see that the resulting set of literals
$P_3$ is a satisfying assignment of $F^* \setminus S^*$, which does
not contain $\neg l^*_1$ nor $\neg l^*_2$. It follows that $S^*$ is
a CS of $(F^*,\{l^*_1\},l^*_2)$ of size at most $k$.

Conversely, let $S^*$ be a CS of $(F^*,\{l^*_1\},l^*_2)$ of size at
most $k$. Let $S$ be a set of clauses of $F$ such that the clause of
index $i$ belongs to $S$ if and only if $C(l_i) \in S^*$ or $C(\neg
l_i) \in S^*$. Clearly $|S| \leq |S^*|$. Let $S^*_2 \subseteq
Clauses(F^*)$ be the set of all clauses $C(l_i)$ and $C(\neg l_i)$
such that the clause of index $i$ belongs to $S$. Since $S^*
\subseteq S^*_2$, we can specify a satisfying assignment $P^*_2$ of
$F^* \setminus S^*_2$ which does not contain $\neg l^*_1$ nor $\neg
l^*_2$.

Let $P$ be a set of literals obtained from $P^*_2$ by removal of all
$l_i$, $\neg l_i$, removal of $l^*_1$ and $l^*_2$, and adding all
the literals $l'$ of $L$ such that $l'$ or $\neg l'$ appear in the
clauses of $F \setminus S$. It is not hard to see that $Var(P)=Var(F
\setminus S)$ and that $P$ does not intersect with $\neg L$.

To observe that $P$ is a satisfying assignment of $F \setminus S$,
note that there is a bijection between the pairs $C(l_i), C(\neg
l_i)$ of clauses of $F^* \setminus S^*_2$ and the clauses of $F
\setminus S$. In particular, each clause of $F \setminus S$ is the
origin of exactly one pair $\{C(l_i),C(\neg l_i)\}$ of $F^*
\setminus S^*_2$ in the form described above and each pair
$\{C(l_i),C(\neg l_i)\}$ of $F^* \setminus S^*_2$ has exactly one
origin in $F \setminus S$.

Now, let $(l' \vee l'')$ be a clause of $F \setminus S$ which is the
origin of $C(l_i)=(t' \vee l_i)$ and $C(\neg l_i)=(\neg l_i \vee
t'')$ of $F^* \setminus S^*_2$, where $l'=t'$ or $t'$ is the result
of replacement of $l'$, $t''$ has the analogous correspondence to
$l''$. By definition of $P^*_2$, either $t' \in P^*_2$ or $t'' \in
P^*_2$. Assume the former. In this case if $l'=t'$ then $l' \in P$.
Otherwise $t' \in \{l^*_1,l^*_2\}$ and, consequently $l' \in L$. By
definition of $P$, $l' \in P$. It can be shown analogously that if
$t'' \in P^*_2$ then $l'' \in P$. It follows that any clause of $F
\setminus S$ is satisfied by $P$.

It follows from the above argumentation that Problem I1 with input
$(F,L,k)$ can be solved by solving the parameterized
\textsc{2-aslasat} problem with input $(F^*,\{l^*_1\},l^*_2,k)$. In
particular, if the output of the \textsc{2-aslasat} problem on
$(F^*,\{l^*_1\},l^*_2, k)$ is a set $S^*$, this set can be
transformed into $S$ as shown above and $S$ can be returned;
otherwise 'NO' is returned. Observe that $|Clauses(F^*)|=O(m)$ and
$|Var(F^*)|=O(m+|Var(F)|)$. Taking into account our note in the
proof of Theorem \ref{2aslasat} that $|Var(F)|=O(m)$,
$|Var(F^*)|=O(m)$. Also note that we may assume that $k<m$ because
otherwise the algorithm can immediately returns $Clauses(F^*)$.

Substituting this data into the runtime of \textsc{2-aslasat}
problem following from Theorem \ref{2aslasat}, we obtain that
problem I1 can be solved in time
$O(5^k*k*m*(m+|\{l^*_1\}|))=O(5^k*k*m^2)$. $\square$

\begin{cclaim} \label{problemi2}
Problem I2 with input $(F,S,k)$ can be solved in time
$O(15^k*k*m^2)$, where, $m=|Clauses(F)|$.
\end{cclaim}

{\bf Proof} We solve Problem I2 by the following algorithm. Explore
all possible subsets $E$ of $S$ of size at most $k$. For the given
set $E$ explore all the sets of literals $L$ obtained by choosing
$l_1$ or $l_2$ for each clause $(l_1 \vee l_2)$ of $S \setminus E$
and creating $L$ as the set of all chosen literals. For all the
resulting pairs $(E,L)$ such that $L$ is non-contradictory, solve
Problem I1 for input $(F^*,L, k-|E|)$ where $F^*=F \setminus S$. If
for at least one pair $(E,L)$ the output is a set $S^*$ then return
$E \cup S^*$. Otherwise return 'NO'. Assume that this algorithm
returns $E \cup S^*$ such that $S^*$ has been obtained for a pair
$(E,L)$. Let $P$ be a satisfying assignment of $F^* \setminus S^*$
which does not intersect with $\neg L$. Observe that $P \cup L$ is
non-contradictory, that $P \cup L$ satisfies all the clauses of
$Clauses(F^* \setminus S^*) \cup (S \setminus E)$ and that
$Clauses(F^* \setminus S^*) \cup (S \setminus E)=Clauses(F \setminus
(S^* \cup E))$. Let $L'$ be a set of literals, one for each variable
of $Var(S \setminus E) \setminus Var(P \cup L)$. Then $P \cup L \cup
L'$ is a satisfying assignment of $F \setminus (S^* \cup E)$, i.e.
the output $(S^* \cup E)$ is valid. Assume that the output of
Problem I1 is 'NO' for all inputs but there is a set $Y \subseteq
Clauses(F)$ such that $|Y| \leq k$ and $F \setminus Y$ is
satisfiable. Let $E=Y \cap S$, $S^*=Y \setminus S$. Let $P$ be a
satisfying assignment of $F \setminus Y$ and let $L$ be a set of
literals obtained by selecting for each clause $C$ of $S \setminus
E$ a literal of $C$ which belongs to $P$. Then the subsets of $P$ on
the variables of $F^* \setminus S^*$ witnesses that $SWRT(F^*
\setminus S^*,L)$ is true that is the output of problem I1 on
$(E,L)$ cannot be 'NO'. This contradiction shows that when the
proposed algorithm returns 'NO' this output is valid, i.e. the
proposed algorithm correctly solves Problem I2.

In order to evaluate the complexity of the proposed algorithm, we
bound the number of considered combinations $(E,L)$. Each clause
$C=(l_1 \vee l_2) \in S$ can be taken to $E$ or $l_1$ can be taken
to $L$ or $l_2$ can be taken to $L$. That is, there are 3
possibilities for each clause, and hence there are at most $3^{k+1}$
possible combinations $(E,L)$. Multiplying $3^{k+1}$ to the runtime
of solving Problem I1 following from Claim \ref{problemi1}, we
obtain the desired runtime for Problem I2. $\square$

Let $(F,k)$ be an instance of \textsc{2-asat} problem without
repeated occurrences of clauses. Let $C_1, \dots, C_m$ be the
clauses of $F$. Let $F_0, \dots, F_m$ be \textsc{2-cnf} formulas
such that $F_0$ is the empty formula and for each $i$ from $1$ to
$m$, $Clauses(F_i)=\{C_1, \dots C_i\}$. We solve $(F,k)$ by the
method of iterative compression \cite{Niederbook}. In particular we
solve the \textsc{2-asat} problems $(F_0,k), \dots (F_m,k)$ in the
given order. For each $(F_i,k)$, the output is either a CS $S_i$ of
$F_i$ of size at most $k$ or 'NO'. If 'NO' is returned for any
$(F_i,k)$, $i \leq m$, then clearly 'NO' can be returned for
$(F,k)$. Clearly, for $(F_0,k)$, $S_0=\emptyset$. It remains to show
how to get $S_i$ from $S_{i-1}$. Let $S'_i=S_i \cup \{C\}$. If
$|S'_i| \leq k$ then $S_i = S'_i$. Otherwise, we solve problem $I2$
with input $(F_i,S'_i,k)$. If the output of this problem is a set
then this set is $S_i$, otherwise the whole iterative compression
procedure returns 'NO'. The correctness of this procedure can be
easily shown by induction on $i$. In follows that \textsc{2-asat}
problem with input $(F,k)=(F_m,k)$ can be solved by at most $m$
applications of an algorithm solving Problem I2. According to Claim
\ref{problemi2}, Problem I2 can be solved in $O(15^k*k*m^2)$, so
\textsc{2-asat} problem with input $(F,k)$ can be solved in
$O(15^k*k*m^3)$.

Finally we show that if $(F,k)$ contains repeated occurrences of
clauses then the \textsc{2-asat} problem remains \textsc{fpt} and
even can be solved in the same runtime. In order to do that, we
transform $F$ into a formula $F^*$ with all clauses being pairwise
distinct and show that $F$ can be made satisfiable by removal of at
most $k$ clauses if and only if $F^*$ can.

Assign each clause of $F$ a unique index from $1$ to $m$. Introduce
new literals $l_1, \dots, l_m$ of distinct variables which do not
intersect with $Var(F)$. Replace the $i$-th clause $(l' \vee l'')$
by two clauses $(l' \vee l_i)$ and $(\neg l_i \vee l'')$. Denote the
resulting formula by $F^*$. It is easy to observe that all the
clauses of $F^*$ are distinct. Let $I$ be the set of indices of
clauses of $F$ such that the formula resulting from their removal is
satisfiable and let $P$ be a satisfying assignment of this resulting
formula. Let $S^*=\{(l' \vee l_i)|i \in I\}$. Clearly, $|S^*|=|I|$.
Observe that $F^* \setminus S^*$ is satisfiable. In particular, for
every pair of clauses $(l' \vee l_i)$ and $(\neg l_i \vee l'')$ at
least one clause is either satisfied by $P$ or belongs to $S^*$.
Hence $F^* \setminus S^*$ can be satisfied by assignment which is
obtained from $P$ by adding for each $i$ either $l_i$ or $\neg l_i$
so that the remaining clauses are satisfied.  Conversely, let $S^*$
be a set of clauses of $F^*$ of size at most $k$ such that $F^*
\setminus S^*$ is satisfiable and let $P^*$ be a satisfying
assignment of $F^* \setminus S^*$. Then for the set of indices $I$
which consists of those $i$-s such that the clause containing $l_i$
or the clause containing $\neg l_i$ belong to $S^*$. Clearly $|S^*|
\geq |I|$. Let $F'$ be the formula obtained from $F$ by removal the
clauses whose indices belong to $I$. Observe that a clause $(l' \vee
l'')$ belongs to $Clauses(F')$ if and only if both $(l' \vee l_i)$
and $(\neg l_i \vee l'')$ belong to $Clauses(F^* \setminus S^*)$. It
follows that either $l'$ or $l''$ belong to $P^*$. Consequently the
subset of $P^*$ consisting of the literals of variables of $F'$ is a
satisfying assignment of $F'$.

The argumentation in the previous paragraph shows that the
\textsc{2-asat} problem with input $(F,k)$ can be solved by solving
the \textsc{2-asat} problem with input $(F^*,k)$. If the output on
$(F^*,k)$ is a set $S^*$ then $S^*$ is transformed into a set of
indices $I$ as shown in the previous paragraph and the multiset of
clauses corresponding to this set of indices is returned. If the
output of the \textsc{2-asat} problem on input $(F^*,k)$ is 'NO'
then the output on input $(F,k)$ is 'NO' as well. To obtain the
desired runtime, note that $F^*$ has $2m$ clauses and $O(m)$
variables and substitute this data to the runtime for
\textsc{2-asat} problem without repeated occurrences of literals.
$\blacksquare$

We conclude the paper by presenting a number of by-products of the
main result. It is noticed in \cite{dagopen} that the parameterized
\textsc{2-asat} problem is \textsc{fpt}-equivalent to the vertex
cover problem parameterized above the prefect matching
(\textsc{vc-pm}). It is shown \cite{RamanISAAC} that the
\textsc{vc-pm} problem is \textsc{fpt}-equivalent to the vertex
cover problem parameterized above the size of a maximum matching and
that the latter problem is \textsc{fpt}-equivalent to a problem of
finding whether at most $k$ vertices can be removed from the given
graph so that the size of the minimum vertex cover of the resulting
graph equals its size of maximum matching. It follows from Theorem
\ref{2asatfpt} that all these problems are \textsc{fpt}.

\subsection*{Acknowledgements}
We thank Venkatesh Raman for pointing out several relevant
references.

\bibliographystyle{plain}

\begin{thebibliography}{20}
\bibitem{Gutinbook}
{\sc J.~Bang-Jensen, G.~Gutin} {\emph Digraphs: Theory, Algorithms
and Applications}, Springer-Verlag, 2001.

\bibitem{ChenWADS2007}
{\sc J.~Chen, Y.~Liu, S.~Lu} {\emph An Improved Parameterized
Algorithm for the Minimum Node Multiway Cut Problem.}, WADS 2007,
pp. 495-506.

\bibitem{ChenWADSFVS}
{\sc J.~Chen, F.~Fomin, Y.~Liu, S.~Lu, Y.~Villanger}, {\em Improved
Algorithms for the Feedback Vertex Set Problems.}, WADS 2007, pp.
422-433.

\bibitem{ChenSTOC}
{\sc J.~Chen, Y.~Liu, S.~Lu, B.~O'Sullivan, I.~Razgon},{\em Directed
Feedback Vertex Set is Fixed-Parameter Tractable}, STOC 2008, to
appear.

\bibitem{ChenKanj2004}
{\sc J.~Chen, I.~Kanj}, {\em Improved Exact Algorithms for MAX-SAT},
Discrete Applied Mathematics, 142(1-3),2004,pp. 17-27.

\bibitem{dagopen}
{\sc E.~Demaine, G.~Gutin, D.~Marx, U.~Stege},{\em Open Problems
from Dagstuhl Seminar 07281}, avaiable electronically at
\url{http://drops.dagstuhl.de/opus/volltexte/2007/1254/pdf/07281.SWM.Paper.1254.pdf}

\bibitem{GroheICALP}
{\sc M.~Grohe, M.~Gr\"uber}, {\em Parameterized Approximability of
the Disjoint Cycle Problem.}, ICALP 2007, pp. 363-374.

\bibitem{GuoSOFFSEM}
{\sc J.~Guo, F.~H\"uffner E.~Kenar, R.~Niedermeier, J.~Uhlmann} {\em
Complexity and Exact Algorithms for Multicut.}, SOFSEM 2006, pp.
303-312.


\bibitem{HuffnerWEA}
{\sc F.~H\"uffner} {\em Algorithm Engineering for Optimal Graph
Bipartization.}, WEA 2005, pp.240-252.

\bibitem{HuffnerSurvey}
{\sc F.~H\"uffner, R.~Niedermeier, S.~Wernicke},{\em Techniques for
Practical Fixed-Parameter Algorithms.}, The Computer Journal, 51(1),
2008, pp. 7-25.


\bibitem{KhotRaman00}
{\sc S.~Khot, V.~Raman}, {\em Parameterized Complexity of Finding
Subgraphs with Hereditary Properties.}, COCOON 2000, pp. 137-147.

\bibitem{RamanECCC}
{\sc M.~Mahajan, V.~Raman} {\em Parametrizing Above Guaranteed
Values: MaxSat and MaxCut}, Electronic Colloquium on Computational
Complexity (ECCC), 4(33),1997.

\bibitem{RamanJALG}
{\sc M.~Mahajan, V.~Raman} {\em Parametrizing Above Guaranteed
Values: MaxSat and MaxCut}, Journal of Algorithms, 31(2), 1999, pp.
335-354.

\bibitem{MarxTCS}
{\sc D. Marx} {\em Parameterized graph separation problems.},
Theoretical Computer Science, 351(3), 2006, pp. 394-406.


\bibitem{RamanISAAC}
{\sc S.~Mishra, V.~Raman, S.~Saurabh, S.~Sikdar, C.R.~Subramanian}
{\em The Complexity of Finding Subgraphs Whose Matching Number
Equals the Vertex Cover Number}, ISAAC 2007, pp. 268-279.


\bibitem{Niederbook}
{\sc R.~Niedermeier}, {\em Invitation to fixed-parameter
algorithms}, Oxford Lecture Series in Mathematics and its
Applications, volume 31, 2006.

\bibitem{Papa}
{\sc C.~Papadimitriou}, {\em Computational Complexity},
Addision-Wesley, 1994.

\bibitem{Reed1}
{\sc B.~Reed, K.~Smith, A.~Vetta}, {\em Finding odd cycle
transversals.}, Operations Research Letters 32(4),2004, pp.299-301.

\bibitem{Slivkins1}
{\sc A.~Slivkins}, {\em Parametrized Tractability of Edge-Disjoint
Paths on DAGs}, ESA 2003, pp.482-493.
\end{thebibliography}

\end{document}